\documentclass[prd,twocolumn,nofootinbib,superscriptaddress,amsmath,amssymb,eqsecnum]{revtex4-1}
\usepackage{amsmath,amssymb, amsthm,amstext}
\usepackage{mathtools}
\usepackage{natbib}
\usepackage{graphicx}
\usepackage{color}
\usepackage{array, enumerate}
\usepackage{bm}
\usepackage{multirow}
\usepackage[breaklinks,colorlinks,citecolor=blue]{hyperref}
\usepackage{braket}
\usepackage{txfonts}
\usepackage[normalem]{ulem}
\usepackage{comment}
\usepackage[utf8]{inputenc}

\usepackage{subfigure}

\def\nn{\nonumber}
\def\be{\begin{equation}}
\def\ee{\end{equation}}

\newcommand\lb{\left(}
\newcommand\rb{\right)}
\newcommand\ba{\begin{eqnarray}}
\newcommand\ea{\end{eqnarray}}
\newcommand\bw{\begin{widetext}}
\newcommand\ew{\end{widetext}}
\renewcommand{\Re}{\operatorname{Re}}

\newcommand{\blue}[1]{\textcolor{black}{#1}}

\definecolor{electricpurple}{rgb}{0.75, 0.0, 1.0}

\begin{document}
\title{Vacuum Spacetime With Multipole Moments: The Minimal Size Conjecture, Black Hole Shadow, and Gravitational Wave Observable}

\author{Shammi Tahura}
\email{sharaban-tahura@uiowa.edu}
\affiliation{University of Guelph, Guelph, Ontario N1G 2W1, Canada}
\affiliation{Perimeter Institute for Theoretical Physics, Ontario, N2L 2Y5, Canada}
\affiliation{Department of Physics and Astronomy, University of Iowa, Iowa City, IA 52242, USA}
\author{Hassan Khalvati}
\affiliation{University of Guelph, Guelph, Ontario N1G 2W1, Canada}
\affiliation{Perimeter Institute for Theoretical Physics, Ontario, N2L 2Y5, Canada}

\author{Huan Yang}
\email{hyang@perimeterinstitute.ca}
\affiliation{University of Guelph, Guelph, Ontario N1G 2W1, Canada}
\affiliation{Perimeter Institute for Theoretical Physics, Ontario, N2L 2Y5, Canada}

\begin{abstract}
In this work, we explicitly construct the vacuum solution of Einstein's equations with prescribed multipole moments. By observing the behavior of the multipole spacetime metric at small distances, we conjecture that for a sufficiently large multipole moment, there is a minimal size below which no object in nature can support such a moment. The examples we have investigated suggest that such minimal size scales as $(M_n)^{1/(n+1)}$ (instead of $(M_n/M)^{1/n}$), where $M$ is the mass and $M_n$ is the $n$th order multipole moment. With the metric of the ``multipole spacetime", we analyze the shape of black hole shadow for various multipole moments and discuss the prospects of constraining the moments from shadow observations. In addition, we discuss the shift of gravitational wave phase with respect to those of the Kerr spacetime, for a test particle moving around an object with this set of multipole moments. These phase shifts are required for the program of mapping out the spacetime multipole moments based on gravitational wave observations of extreme mass-ratio inspirals.
\end{abstract}
\maketitle

\section{Introduction}
Black holes are the most compact astrophysical objects in the universe as predicted by the theory of General Relativity. Since the direct observation of gravitational waves in 2015, various properties of black holes (e.g. the ringdown tests \cite{LIGOScientific:2020tif,LVK2021gr,Berti:2016lat,Yang:2017zxs,Isi:2019aib,Berti:2018vdi}) have been examined in the strong-gravity regime. In recent years the observation of black hole image (``shadow") using very long baseline interferometry (VLBI)~\cite{EventHorizonTelescope:2019dse,EventHorizonTelescope:2022wkp} provides an alternative probe to the spacetime region within several gravitational radii. In order to further facilitate the test of black holes, many options of black hole mimickers have been discussed in the literature, including ultra-compact objects~\cite{Vincent:2015xta,Olivares:2018abq,Rosa:2023qcv,Cardoso:2019rvt,Raposo:2018xkf}, gravastar/ AdS bubbles \cite{Pani:2009ss,Danielsson:2021ykm,Yang:2022gic,Mazur:2001fv}, wormholes~\cite{Tsukamoto:2012xs,Tsukamoto:2016zdu,Mazza:2021rgq,Poisson:1995sv}, etc.  It is plausible that these black hole mimickers have a different set of multipole moments from a Kerr black hole with the same mass and spin. As a result, sometimes it is useful to define a test of black hole mimickers as measuring the multipole moments of the spacetime as initiated in \cite{Ryan:1995wh,Ryan:1997hg}, although theoretically it is possible to have a non-black hole object with exactly the same multipole moments as Kerr \cite{Bonga:2021ouq}.

In Newtonian gravity it is straightforward to obtain the multipole moments from the  decomposition of  gravitational field. The Newtonian gravitational field is characterized by its multipole moments in the sense that given a set of multipole moments, the gravitational potential can be constructed uniquely. On the other hand, as the Einstein equations are nonlinear, the same statement is nontrivial in General Relativity (GR). In the general-relativistic context, a definition of multipole moment has been given by Geroch for a vacuum static spacetime~\cite{Geroch:1970cd} and later by Hansen for stationary spacetimes~\cite{Hansen:1974zz}. Geroch and Hansen's definition of multipole moments are given in terms of conformal compactification of trajectories of time translation Killing vectors. The moments are totally symmetric trace-free (STF) tensors at spatial infinity constructed from derivatives of the norm and twist of the spacetime. The definition is mathematically elegant and manifestly coordinate-independent.

Another definition was given by Kip Thorne for slowly changing fields suitable for studying gravitational waves (GWs) in the post-Newtonian context~\cite{Thorne:1980ru}. Thorne's method adopts the physical metric in De Donder gauge in ACMC (asymptotically Cartesian and mass-centered) coordinates. The metric is expanded in inverse of radial coordinates in terms of two sets of mass and current multipole moments. The equivalence of the two definitions was proven for slowly evolving stationary systems modulo some normalization constant~\cite{gursel1983}. Geroch-Hansen and Thorne Multipole moments have also been defined in some modified or alternative theories of gravity~\cite{Sopuerta:2009iy,Pappas:2014gca,Cano:2022wwo}.

As we are interested in the strong-field gravity regime where the post-Newtonian approach may not be applicable, we will consider the Geroch-Hansen multipole moments in this paper. For an axisymmetric stationary system, the Geroch-Hansen multipole moments can be computed relatively easily via Ernst formalism: an expansion of complex Ernst potential along the symmetry axis at spatial infinity~\cite{Ernst:1967wx}. Following such an approach, an algorithm was provided in Ref.~\cite{Fodor1989} for computing multipole moments of stationary axisymmetric systems in vacuum GR. Using a similar approach, multipole moments were also computed for radiating systems~\cite{Ryan:1995wh} and electrovacuum spacetimes~\cite{Fodor:2020fnq,Sotiriou:2004ud}. In particular, recently Fodor \emph{et al.} (Ref.~\cite{Fodor:2020fnq}) provided an efficient algorithm for computing gravitational and electromagnetic multipole moments in stationary electrovacuum spacetime. They implement the tools of complex null vector fields and the idea of leading order parts of functions introduced by B\"{a}ckdahl and Herberthson~\cite{Backdahl:2005be,Backdahl:2006ed}, which simplifies the computation of STF part of a tensor with reduced number of variables.

Given a set of Geroch-Hansen moment, it has been shown that the vacuum regime of the exterior spacetime is indeed uniquely determined \cite{Geroch:1970cd}, similar to the Newtonian case. In addition, the mathematical procedure described in \cite{Fodor:2020fnq} may be reversed to construct the exterior metric associated with the given moments. In this work, we explicitly carry out the procedure to construct the ``multipole spacetime" - the vacuum spacetime with prescribed moments. Namely, we use the algorithm in Ref.~\cite{Fodor:2020fnq} to compute the coefficients of power series expansion of complex Ernst potential recursively from multipole moments. The norm and twist of the spacetime then can be extracted from the Ernst potential and any other metric functions can be computed from these two by directly integrating the Ernst equations.  

We apply the multipole spacetime for two main objectives. First, it helps to study properties of the underlying source, i.e., the source size. Second, because of the one-to-one correspondence between the spacetime and the moments, we can discuss possible gravitational wave and electromagnetic observables in the strong-gravity regime that can be used to infer the spacetime multipole moments. Note that the original proposal in \cite{Ryan:1995wh,Ryan:1997hg}  is only valid in the asymptotic region.

In the first objective, we are particularly interested in the connection between the size of the source and the associated moments. For the monopole moment, i.e., the mass $M$, Thorne conjectured that the size of the circumference enclosing a source cannot be smaller than $\mathcal{O}(1) M$ (as indicated from the ``Hoop Conjecture").
With a similar motivation, it is instructive to ask the question that, {\it are the generic mass and current moments $M_\ell, S_\ell$ related to the minimal size of the source as well}?
In Newtonian gravity, it is obvious that such a relation exists since the field multipole moments of the gravitational potential are identical to the source multipole moments. The scaling of the multipole moments is such that the $n$-th order multipole moment scales as $M L^n$ where $L$ is the characteristic size of the source. In the post-Newtonian regime, Thorne's multipole moments can be written as integrals of some effective stress-energy tensor of the source. The mass multipole moments in the case of a slow-moving source in weak-field in Thorne formalism also generically scale as $M L^n$(see Eq.~5.21 of~\cite{Thorne:1980ru}). 
It is not obvious that the same scaling will hold in the strong-gravity regime when the effect of curvature is non-negligible.  Nevertheless, it is physical to expect that a bound source cannot produce arbitrarily large field multipole moments even in strong-field gravity. As a result, we expect that there is a \emph{minimal size} source that can support  a particular set of multipole moments. In the large moment limit, this size limit may satisfy certain scaling law with the moments.

The multipole spacetime, as a solution of the vacuum Einstein's equations, is constructed based on the moments extracted at spatial infinity and then extended towards smaller radius. The formalism in \cite{Fodor:2020fnq} allows the metric to be written in powers of $1/r$, so that by including more terms the power-law expansion asymptotes the true solution for generic radius. However, we find that the multipole spacetime cannot be extended all the way to the origin at $r=0$ -- the solution always hits a curvature singularity at a finite radius (depending on the angular directions). Therefore, no matter what kind of sources is generating the multipole moments at infinity, the source size cannot be smaller than the radius where we encounter the singularity, i.e., this radius can serve as a lower bound on the size of the source. For  a static axisymmetric spacetime with a large multipole moment of order $n$, such an approach suggests that the minimal size of a source scales as $M_n^{1/(n+1)}$, which is different from the intuition in the Newtonian regime: $(M_n/M)^{1/n}$. It is plausible that such scaling law applies for general stationary spacetimes, but so far we have not found an explicit construction of a source that can generate a multipole moment $M_n$ with size $\sim M_n^{1/(n+1)}$. On the other hand, we have also studied the regime that the multipole moments  only  differ from the Kerr values by a small magnitude. The corresponding location of singularity is close to the Kerr horizon radius.

The multipole spacetime metric for the latter case is particularly useful as we are interested in how well various observables can be used to measure the difference between the spacetime multipoles from the Kerr values. The sensitivity  characterizes our ability of probing/constraining black hole mimickers. In this work we analyze two commonly discussed experiments in the literature: black hole shadow/critical curve measurement with VLBI and gravitational wave measurement on extreme mass-ratio inspirals (EMRIs).

We construct the multipole spacetime with weakly perturbed quadrupole moment $M_2$ from the Kerr value while keeping other moments unchanged. The spacetime critical curve is then computed with respect to this spacetime metric. In addition, since the black hole spin, mass and the observer's inclination angle are not known in priori, we vary these parameters to generate Kerr critical curves that best mimic the one of the multipole spacetime. The resulting relative mismatch (see the definition in Sec.~\ref{sec:4}) is around $\delta M_2/(40 M^3)$. Notice that if $\delta M_2 $ is large (see Fig.~\ref{fig:rBLvsdeltam2_curv}), the minimal size of the source might exceeds the size of the light ring, so that there is no meaningful critical curve of the spacetime. For cases where the critical curve still lies within the light ring, the mismatch obtained here is likely not resolvable by the next-generation Event Horizon telescope, which is expected to sample at most the $n = 1$ light ring with its longest baselines (at 345 GHz).

Secondly, we explore the prospect of probing the spacetime multipole moments with GWs from extreme-mass-ratio inspirals (EMRIs). EMRIs  are mainly produced by scattering in the nuclear cluster and disk-assisted migration \cite{Babak:2017tow,Pan:2021ksp,Pan:2021oob}, and they are one of the main sources of future space-based GW observatory, e.g. LISA, Taiji and Tianqin~\cite{Audley:2017drz,Hu:2017mde,TianQin:2020hid}. They are ideal for  probing the spacetime as the stellar-mass compact object is generally in-band for $10^4-10^5$ cycles during the observation period, so that small variation in the spacetime geometry may lead to accumulated phase shift over many cycles. For this study we focus on the non-resonant part of the waveform, where the discussion of the general EMRI resonant dynamics in a perturbed Kerr spacetime can be found in \cite{Pan:2023wau}. For a sample system with $10^6 M_{\odot}$ host black hole and $10 M_\odot$ secondary black hole, we find that a four-year observation can constrain $\delta M_2$ to the $10^{-4}$ level.

The rest of the paper is organized in the following manner. Sec.~\ref{sec:2} gives the necessary definitions of metric and multipole moments and develops the framework for computing metric components from multipole moments. Sec.~\ref{sec:conv_method} discusses the methodology we implement to estimate the the size of a source from properties of metric and curvature. Sec.~\ref{sec:3A} focuses on computing the location of divergence of static axisymmetric metric and curvature with a large multipole moment, while Sec.~\ref{sec:3B} has a similar target for a stationary spacetime with a small deviation from Kerr quadrupole. In Sec.~\ref{sec:4} we compute BH shadows for the metric in Sec.~\ref{sec:3B} and find the mismatch between such shadows and Kerr BH shadows. In Sec.~\ref{sec:5}, we derive the EMRI waveforms for the same metric as in Sec.~\ref{sec:3B}. Finally, we present concluding remarks in Sec.~\ref{sec:6}. Throughout the paper, we use the geometrized unit system $G=c=1$.
\section{Stationary Spacetime with Multipole Moments}\label{sec:2}
In this section, we  briefly present the formalism of computing multipole moments of stationary spacetimes following the approach given by Hansen~\cite{Hansen:1974zz}. We also discuss how to reconstruct the metric of a spacetime from a set of multipole moments in axisymmetric spacetimes, with the formalism discussed in \cite{Fodor:2020fnq}. Hansen's approach is based on quantities defined at spatial infinity via conformal compactification of a three manifold of timelike Killing vector trajectories, {i.e., the manifold to which the Killing vector field is tangent everywhere~\cite{Geroch:1970nt,Hansen:1974zz}}.

\subsection{Definition of Geroch-Hansen multipole moments}
Let us consider a four dimensional manifold $\mathcal M$ with a metric $g_{ab}$ of signature \textcolor{black}{$(-, +, +, +)$} that has a timelike Killing vector $\xi^{a}$. The induced metric $h_{ab}$ on the 3 dimensional submanifold $V$ of the trajectories of $\xi^a$ is positive definite and is related to the full spacetime metric in the following way:
\be\label{eq:gtoh}
h_{ab}=\lambda\, g_{ab}+\xi_{a}\xi_{b}\,.
\ee
$\lambda=-\xi^{a}\xi_{a}$ is the norm of the Killing vector which  is the analog of Newtonian potential for static spacetimes~\cite{Geroch:1970cd}. Let us also denote the covariant derivative of the metric $h_{ab}$ as $D_a$. $(h_{ab},V)$ is defined to be asymptotically flat  by requiring that a manifold $\tilde V$ with a metric $\tilde h_{ab}$ exists such that\cite{Penrose:1965am,Geroch:1970cd}
\begin{enumerate}
\item $\tilde V=V\cup\Lambda$, where $\Lambda$ is a single point\,,
\item $\tilde h_{ab}=\Omega^2 h_{ab}$ is a smooth metric on $\tilde V$\,,
\item $\Omega\,\big|_{\Lambda}=0,\,\tilde D_a \Omega\,\big|_{\Lambda}=0,\,\tilde D_a \tilde D_b\Omega\,\big|_{\Lambda}=2\tilde h_{ab}\big|_{\Lambda}$.
\end{enumerate}
Here $\Omega$ is the conformal factor and the point $\Lambda$ corresponds to the spacelike infinity. $\tilde D_a$ is the covariant derivative of the metric $\tilde h_{ab}$.

Let us now define a complex function $\phi$ on V such that $\tilde \phi=\Omega^{-1/2}\phi$ extends smoothly to $\Lambda$ on $\tilde V$. The multipole moments are then  defined with certain tensorial quantities at $\Lambda$ containing $\tilde \phi$ and its derivatives. Let us define $P^{(0)}=\tilde \phi$ {and a set of multi-index tensors $P^{(1)}_{\tilde a_1},\,P^{(2)}_{\tilde a_1\tilde a_2},\,P^{(3)}_{\tilde a_1\tilde a_2\tilde a_3}\ldots$ recursively as}
\be\label{eq:multipole-def1}
P^{(n)}_{\tilde a_1\,\cdots \tilde a_n}=C\left[\tilde D_{\tilde a_1}P_{\tilde a_2\,\cdots \tilde a_n}^{(n-1)}-\frac{1}{2}(n-1)(2n-3)\tilde R_{\tilde a_1\tilde a_2}P^{(n-2)}_{\tilde a_3\,\cdots \tilde a_n}\right]\,.
\ee
{Here we have chosen a set of coordinates $x^{\tilde{a}}$ on the manifold $\tilde V$. $\tilde R_{\tilde a \tilde b}$ is the Ricci tensor of the metric $\tilde h_{\tilde a\tilde b}$ and the operator $C[\cdots]$ takes the symmetric trace-free projection of its argument.}The $n$-th order multipole moment of the spacetime is then the tensor $P^{(n)}_{\tilde a_1\,\cdots \tilde a_n}$ evaluated at $\Lambda$~\cite{Hansen:1974zz,Geroch:1970cd}:
\be\label{eq:multipole-def2}
M^{(n)}_{\tilde a_1\,\cdots \tilde a_n}=P^{(n)}_{\tilde a_1\,\cdots \tilde a_n}\big|_{\Lambda}\,.
\ee
\subsection{Axisymmetric spacetimes with prescribed multipole moments}\label{sec:2b}
The metric of a stationary axisymmetric vacuum spacetime is suitably expressed in Weyl-Papapetrou coordinates. The field equations are also rather simplified in such coordinates~\cite{papapetrou1953}. We consider the Weyl-Papapetrou coordinate $(t,\rho,z,\phi)$ with z being the axis of symmetry:
\begin{equation}\label{eq:gstationary}
d s^2=-f(d t-\omega d\varphi)^2+\frac{1}{f}\left[e^{2 \gamma}\lb d\rho^2+\mathrm{d} z^2\rb+\rho^2 d\varphi^2\right]\,.
\end{equation}
The metric functions $f$, $\gamma$, and $\omega$ depend on $\rho$ and z only. $\omega$ characterizes the rotation of the spacetime, so that setting $\omega=0$ leads to a static spacetime (which we will use when we first introduce the minimal size conjecture). Comparing Eq.~\eqref{eq:gstationary} to Eq.~\eqref{eq:gtoh} we can find that $\lambda=f$ is the norm and the metric on the submanifold V is given by $h_{ab}\equiv \mathrm{Diag}[e^{2\gamma},\,e^{2\gamma},\,\rho^2]$.

A complex Ernst potential for the spacetime above is defined as $\mathcal E=f+\mathrm{i}\chi$~\cite{Ernst:1967wx}, where
\be\label{eq:omegawp}
\partial_{\rho}\chi=-\rho^{-1}f^2\partial_z\omega, \quad \partial_{z}\chi=\rho^{-1}f^2\partial_\rho\omega\,.
\ee
Integrating the above equation, one can obtain $\omega$ from $f$. From Einstein's equations, the following set of equations can be obtained for $\gamma$, which can be solved hierarchically from metric functions $\omega$, and $f$~\cite{Griffiths:2009dfa}:
\ba\label{eq:gamma1}
\partial_\rho \gamma&=&\frac{1}{4}\rho f^{-2}\left[(\partial_\rho f)^2-(\partial_z f)^2\right]-\frac{1}{4}\rho^{-1}f^2\left[(\partial_\rho \chi)^2-(\partial_z \chi)^2\right]\,,\nn \\ \\
\label{eq:gamma2}\partial_z\gamma&=&\frac{1}{2}\rho f^{-2}\partial_\rho f\partial_z f-\frac{1}{2}\rho^{-1}f^2\partial_\rho\chi\partial_z\chi\,.
\ea
Let us define a new potential $\xi=(1-\mathcal E)/(1+\mathcal E)$, then from Einstein's equations, the so-called Ernst equation is obtained in the following form:
\be 
(\xi \bar \xi-1)D^2 \xi=2\bar \xi D^a\xi D_a \xi\,,
\ee
where an overhead bar denotes the complex conjugate of a quantity and $D^2=D^aD_a$, where $D_a$ is the covariant derivative compatible with $h_{ab}$.

By choosing a conformal factor of $\Omega=1/r^2$ and a new set of coordinates $\tilde{\rho}=\frac{\rho}{r^2}$ and $\tilde{z}=\frac{z}{r^2}$, starting from the metric $h_{ab}$ in \eqref{eq:gstationary}, a conformally transformed metric $\tilde h_{\tilde a\tilde b}=\tilde r^4 h_{\tilde a \tilde b}$ can be obtained, where $\tilde r^2=\tilde \rho^2+\tilde z^2$. The spatial infinity in the coordinate system $(\tilde \rho,\tilde z,\phi)$ has coordinate values  $\tilde \rho =\tilde z=0$. In addition, the Ernst equation can be expressed in terms of the conformally rescaled potential $\tilde \xi=\Omega^{-1/2}\xi$ on $\tilde V$ as:
\be\label{eq:ernst-conf}
(\tilde r^2\tilde \xi \bar{\tilde \xi}-1)\tilde D^2 \tilde \xi=2\bar{\tilde \xi}\tilde D^{\tilde a}(\tilde r \tilde \xi)\tilde D_{\tilde a}(\tilde r \tilde \xi)\,.
\ee

Let us now present  the definition of multipole moments of the spacetime described by  Eq.~\eqref{eq:gstationary}. We will choose $\tilde \phi=\tilde \xi$ for the multipole moments defined in Eq.~\eqref{eq:multipole-def1}. For a stationary axisymmetric spacetime, multipole moments take a simple form in terms of a scalar potential $M_n$ and products of unit vectors along the symmetry axis~\cite{Hansen:1974zz}. In our case, they are ~\cite{Fodor:2020fnq}: 
\be
M^{(n)}_{\tilde a_1\,\cdots \tilde a_n}=\left.\frac{2n!}{2^n n!}\,M_n C[n_{\tilde a_1}\cdots n_{\tilde a_n}]\right |_{\Lambda}\,.
\ee
Consequently, we have
\be\label{eq:multipole-def3}
M_n=\left.\frac{1}{n!} M_{\tilde{a}_1 \ldots \tilde{a}_n}^{(n)} n^{\tilde{a}_1}\,\ldots n^{\tilde{a}_n}\right|_{\Lambda}=\frac{1}{n!} M_{\tilde{z}\,\ldots \tilde{z}}^{(n)}
\ee

For a stationary spacetime, multipole moments can be calculated in terms of coefficients of expansion of $\tilde \xi$ on the symmetry axis. Let us adopt an expansion of $\tilde{\xi}=\sum_{k=0,l=0}^{\infty} a_{k l} \tilde{\rho}^k \tilde{z}^l$  which in general becomes $\tilde{\xi}=\sum_{n=0}^{\infty}m_n\tilde z^n$ on the symmetry axis. Namely, the coefficients satisfy $a_{0l}=m_l$. Plugging these expansions into the Ernst equation in Eq.~\eqref{eq:ernst-conf},  one can obtain a recursive relation that can be used to generate all $a_{kl}$ in terms of $m_n$ \cite{Fodor:2020fnq}:
\bw
\ba\label{eq:recursion}
(r+2)^2 a_{r+2, s}=-(s+2)(s+1) a_{r, s+2} +\sum_{\substack{k+m+p=r \\l+n+q=s}} & a_{k l} \bar{a}_{m n} \left[a_{p q}\left(p^2+q^2-2 p-3 q-2 k-2 l-2 p k-2 q l-2\right)\right.\nn
\\&\left.+a_{p+2,q-2}(p+2)(p+2-2k)+a_{p-2,q+2}(q+2)(q+1-2l)\right]\,.
\ea
\ew
\textcolor{black}{Note that in Ref.~\cite{Fodor:2020fnq}, above recursion relation also includes terms related to electromagnetic multipole moments (see Eq.~[79] and Eq.~[80] of Ref.~\cite{Fodor:2020fnq}). The terms related to electromagnetic moments correct some errors of Ref.~\cite{CHoenselaers_1990}, but the gravitational multipole moments seem to agree with those in Ref.~\cite{CHoenselaers_1990} and also in Ref.~\cite{Fodor1989}.}

\textcolor{black}{With above equation in Eq.~\ref{eq:recursion}, $\tilde \xi$ can be computed everywhere on $\tilde V$. Furthermore, using $\tilde \phi=\tilde\xi$, scalar multipole moments $M_n$ can also be obtained. To do so, one first computes $\tilde \xi$ as a function of $m_n$ using the recursion in Eq.~\ref{eq:recursion}. Then, one can compute the derivatives of $\tilde\xi$ and the Ricci tensors $R_{\tilde a \tilde b}$ and use them to Eqs.~\eqref{eq:multipole-def1}-\eqref{eq:multipole-def2} and Eq.~\eqref{eq:multipole-def3} to compute scalar multipole moments as a function of $m_n$. To facilitate the calculation, Ref.~\cite{Fodor:2020fnq} implemented the concept of the leading order part of a function and introduced complex null vectors following Refs.~\cite{Backdahl:2005be,Backdahl:2006ed} to make the procedure of taking symmetric trace-free projection easier. Such techniques allow one to derive higher-order multipole moments in a simplified manner. For example, the moments $M_n$ are evaluated and expressed in terms of $m_n$ up to $n=6$ in Ref.~\cite{Fodor:2020fnq}. Note that for a Kerr spacetime, such moments are simply $M_n=m_n=M(\mathrm{i}a)^n$ with $M$ and $a$ denoting the mass and  spin parameter $J/M$, respectively~\cite{Sotiriou:2004ud,Fodor:2020fnq}}.

\textcolor{black}{In this work, we will focus on a set of prescribed multipole moments and compute the metric functions from the moments. To do so, the procedure above needs to be reversed. First, however, one has to compute the moments $M_n$ as a function of coefficients of Ernst potential $m_n$ using the procedure described above up to a certain order (a Mathematica notebook is provided in Ref.~\cite{Fodor:2020fnq} that includes the computation of moments as a function of $m_n$). Then, solving for $m_n$ in terms of $M_n$, one can use Eq.~\eqref{eq:recursion} to compute $\tilde\xi$ as a function of $M_n$ and evaluate $\xi=\tilde \xi/r$. The set of moments we assume are identical to those of the Kerr spacetime moments $M_n=M(\mathrm{i}a)^n$ except for $M_2$. In other words, $M_2$ has a small deviation from that of Kerr, namely, $M_2=-Ma^2+\delta M_2$. From the real and imaginary parts of $\xi$ we can then obtain $f$ and $\chi$, respectively, which we use to compute $\omega$ from Eq.~\eqref{eq:omegawp}. Finally, $\gamma$ is obtained by using $f$ and $\omega$ computed in Eq.~\eqref{eq:gamma1} or Eq.~\eqref{eq:gamma2}.} To avoid conical singularity on the symmetry axis, we impose the boundary conditions that $\omega$ and $\gamma$ vanish on the symmetry axis. In order to guarantee the asymptotic flatness, $\gamma$ and $\omega$ must vanish as $\rho\rightarrow\infty$ or $z\rightarrow\infty$. Because the metric functions are all computed in a power-law expansion form of $1/r$,  the level of accuracy of the metric (or the order in $1/r$ up to which Einstein equations are satisfied) depends on up to which order in $1/r$ the potential $\xi$ is computed. Denoting $n$ as the highest order of multipole moments considered, $\xi$ is accurate up to $(1/r)^{n+1}$, $f$ is accurate up to $(1/r)^{n+1}$, $\omega$ is accurate up to $(1/r)^n$, and $\gamma$ is accurate up to $(1/r)^{n+1}$.

\section{Convergence Radius and The Minimal Size Conjecture}
In this section, we define the minimal size conjecture and discuss how to compute the minimal size of a source that generates certain multipole moments at infinity. In Sec.~\ref{sec:conv_method}, we present various methods we adopt to compute the minimal size of the source. In Sec.~\ref{sec:3A}, using the metric computed with the Ernst formalism, we derive the minimal size in the case of a static axisymmetric source, in the large multipole moment limit. In Sec.~\ref{sec:3B}, we consider a stationary axisymmetric spacetime that is weakly perturbed from Kerr. With a small deviation to the Kerr quadrupole moment, we analyze the associated minimal size of the source.

Given the mass of an object, the minimal size (or maximum compactness) of the object is set by the limit that the object is a black hole. It is then an interesting question that, if we know the multipole moments of the object, do they provide additional constraint of the size of the object as well? Beside the theoretical interests, this question also has its own practical applications because if the size of black hole mimicker is larger than the light-ring size or the radius of Inner Most Stable Orbit, it will significantly influence a VLBI measurement on the spacetime critical curve and gravitational wave measurement using EMRIs.

In Newtonian gravity, there is a correlation between a multipole moment and the size of a source creating such a moment. This is due to the fact that the field multipole moments and source multipole moments are identical in Newtonian gravity for isolated objects~\cite{poisson_will_2014}. A multipole moment decomposition of Newtonian gravitational potential outside a source in spherical polar coordinates $(r,\theta,\phi)$ is given by

\be\label{eq:UN}
U_N=\sum_{\ell,m}\frac{4\pi}{2\ell+1} I_{\ell m}(t) \frac{Y_{\ell m}(\theta,\phi)}{r^{\ell+1}}\,.
\ee
The multipole moments of the mass distribution of the source $I_{\ell m}(t)$ is
\be\label{eq:Nmultiple}
I_{\ell m}=\int \rho(t,\vec r) r^{\ell} Y_{\ell m}(\theta,\phi)d^3r\,,
\ee
where $\rho (t,\vec r)$ is the mass density. Here multipoles $I_{\ell m}$ scale as $M L^{\ell}$ with $L$ denoting the characteristic size of the source. Then, one is motivated to assume that the minimum size of a source creating such a $\ell$-th order moment should scale as $(I_{\ell m}/M)^{1/\ell}$. This scaling, however, is not guaranteed to hold in strong-field. In fact, based on the analysis of static spacetimes in Sec.~\ref{sec:3A}, we conjecture that {\it the minimal size of a compact source with a given multipole moment $M_n$, when it is sufficiently large, should scale as $(M_n)^{1/(1+n)}$}. Notice that theoretically  the  central object does not necessarily have to be a star-like body, it may also be a surface that is attached to another patch of spacetime like a wormhole, although the stability may be another issue \cite{Yang:2022gic}.

Notice that the multipole spacetime metric discussed in Sec.~\ref{sec:2b} is a series expansion in $1/r$. The expression is completely regular near the spatial infinity, but as the $r$ decreases, it may fail to converge at finite radius. We refer this radius as the ``convergence radius" and check it actually corresponds to a curvature singularity instead of coordinate artifact. Once this is confirmed, the convergence radius actually sets a lower bound on the size of the object, because the vaccum spacetime that is smooth at spatial infinity cannot be regularly extended beyond this point, assuming the vacuum Einstein's equation is still valid.

\subsection{Methodology}\label{sec:conv_method}
Let us consider the Following Taylor series,
\be\label{eq:TaylorSeries}
F(x)=\sum_{n=0}^{\infty}a_n x^n\,,
\ee
where the coefficients $a_n$ are real and independent of the variable $x$. For such a series, it is not clear what will be the most efficient way of determining the convergence radius, as the asymptotic behavior of $a_n$ can be rather complicated. Various converge tests exist, and for a power series, usually, the tests provide an interval of convergence for the variable $x$~\cite{Boas:913305}. In realistic implementations, we only have  a finite number of terms in the power-law expansion because of the computational cost, which limits the performance of some of the more sophisticated methods. The latter point is particularly relevant for the analysis in Sec.~\ref{sec:3B}. We adopt three different methods for obtaining the convergence radius of the spacetime metric, which we present below. For the first two tests (ratio test and root test) readers may refer to Ref.~\cite{ARFKEN20131} or Chapter~1 of Ref.~\cite{Boas:913305}
\subsubsection{Ratio Test}
Ratio test (also known as the d’Alembert ratio test or Cauchy ratio test) is a measure of absolute convergence of the series. Absolute convergence means that if we replace the terms in the series with their absolute values the resulting series is convergent. Note that the actual series may still be convergent if it is not absolutely convergent, and in that case the series is called conditionally convergent. For the series in Eq.~\eqref{eq:TaylorSeries}, the ratio test states that the convergence radius is $R=\lim\limits_{n \to \infty} \left|\frac{a_n}{a_{n+1}}\right|$. The series converges when $|x|<R$ and diverges if $|x|>R$, but the ratio test cannot tell us if the series is convergent or divergent at the boundary $x=R$, so one requires other methods to check convergence on the boundary. Schwarzshild metric is a nice example where ratio test works well. If we expand the $g_{tt}$ component of Schwarzschild metric in Schwarzschild coordinates in terms of inverse distance and use the ratio test, we find the radius of convergence as $2M$, which is the location of the horizon and $g_{tt}$ does blow up on the horizon. However, ratio test for a power series is not particularly useful if the ratios oscillate in the large $n$ limit, which we will encounter in Sec.~\ref{sec:3A}.

\subsubsection{Root Test:}
For the series in Eq.~\eqref{eq:TaylorSeries}, the radius of convergence is $R=1/\lb\limsup\limits_{n\to\infty}\left|a_n\right|^{1/n}\rb$ and the series converges in the interval $-R<x<R$. This test is known as the Cauchy root test or simply root test. Similar to the ratio test, root test gives absolute convergence and does not give information of convergence on the boundary $|x|=R$.
\subsubsection{Upper-Bound Test:}
In an actual physics problem, the coefficients $a_n$ are functions of physical parameters, and most likely, what is relevant is how the convergence radius varies if one changes a physical parameter. An upper-bound test is useful in that regard. First, we set an upper bound for $F$ in Eq.~\eqref{eq:TaylorSeries}, which may have only a finite number of terms on right-hand side. Let us consider that all coefficients are functions of one single physical parameter $b$. Then choosing a value for the physical parameter, one can solve for $x$ that gives that upper bound. Varying $b$ and repeating the procedure then we can find relation between $x$ and $b$ so that F is fixed. If the relation does not change by changing the upper bound on F to various high values, we can assume that the same relation will hold if F could be set to infinity, and so the convergence radius will have a similar scaling with respect to $b$.

\subsection{Static Spacetimes and the Minimal Size Conjecture}\label{sec:3A}
In order to investigate the convergence radius of a multipole spacetime, in this section we assume a static axisymmetric spacetime by setting $\omega=0$, which allows us to obtain semi-analytical results. Let us choose a new set of coordinates  on the submanifold $V$ characterized by $(r,\theta)$ by defining $\rho=r \sin{\theta}$ and $z=r \cos{\theta}$ (note that these coordinates are not necessarily the same as the isotropic spherical polar coordinates in Eq.~\eqref{eq:UN} and Eq.~\eqref{eq:Nmultiple}). Let us also define $f=e^{2U}$.  The metric in Eq.~\eqref{eq:gstationary} can be rewritten as
\be\label{eq:met_static}
ds^2=-e^{2U} dt^2+e^{-2U}\left[e^{2\gamma}(dr^2+r^2d\theta^2)+r^2\sin^2\theta d\phi^2\right]\,.
\ee
For small $U$, we expect that $e^{2U}\simeq 1+2U$ where $U$ is the analog of Newton's potential. In our case, we will see that $U$ can be large as we work in strong-field scenario. The function $U$ satisfies the Laplace equation $D^2U=0$, for which the exact solution is known in terms of Legendre polynomials:
\be\label{eq:U}
U=-\sum_{n=0}^{\infty}A_n r^{-(n+1)}P_n(\cos{\theta})\,.
\ee
Notice that $P_l$ here are the Legendre polynomials (not to be confused with multipole moments), and the coefficients $A_n$ are functions of multipole moments. Corresponding $\gamma$ is~\cite{Griffiths:2009dfa}
\be\label{eq:gamma}
\gamma=-\sum_{l=0}^{\infty}\sum_{m=0}^{\infty}A_l A_m r^{-(l+m+2)}\frac{(l+1)(m+1)}{(l+m+2)}\lb P_lP_m-P_{l+1}P_{m+1}\rb\,.
\ee

For simplicity, we will keep both $M$ and $M_2$ moments and set all other moments to be zero. Implementing the algorithm given Eq~\eqref{eq:recursion} we can find $f$ for the static metric from which we compute $U$ as $U=(1/2)\log{f}$. Expanding $U$ in terms of $1/r$ and matching with Eq.~\eqref{eq:U}, one can extract $A_n$. In the case there are only nonzero $M$ and $M_2$ moments, we find that $A_n=0$ for odd $n$.

Now let us further assume that $M_2 \gg M^3$, which basically says that the scale defining the quadrupole moment is much bigger than that of the monopole moment.
Consequently, at each order in $1/r$, the term with the highest power in $M_2$ dominates. Keeping only such terms at each order, we find a power series expansion of $U$ in $1/r$ that can  be rearranged in the following closed-form expression:
\ba\label{eq:USeries}
U&=&-\sum_{n=0}^{\infty} \left[\mathcal{M}_{n} P_{6n+2}(\cos{\theta})\,\kappa^{6n+3}\right.\nonumber\\ &&\left.
+\frac{M}{M_2^{1/3}}\frac{(2 n+1) (5 n+1) (10 n+3)}{3 (3 n+1) (6 n+1)}\mathcal M_{n}P_{6n}(\cos{\theta})\,\kappa^{6n+1}\right.\nonumber\\&&\left.
+\frac{M^2}{M_2^{2/3}}\frac{(2 n+1) (3 n+2) (5 n+8)}{2 (10 n+7)}\mathcal{M}_{n}P_{6n+4}(\cos{\theta})\,\kappa^{6n+5}\right]\,,\nonumber\\
\ea
with
\be
\mathcal M_{n}=\frac{2^{-2 n-1} 15^{n+1} (6 n+2)!}{(2 n)\text{!!} (10 n+5)\text{!!}}\,,
\ee
\textcolor{black}{where we defined a new variable $\kappa=M_2^{1/3}/r$. Monopole-quadruple solutions of static axisymmetric spacetime was previously studied in Ref.~\cite{Backdahl:2005uz}. By keeping the leading order term in $M_2$ at each order in Eq.~(26) of Ref.~\cite{Backdahl:2005uz}, one can obtain above series with a rescaling of of $M_2\rightarrow 3M_2$. The first series in Eq~\eqref{eq:USeries} corresponds to the pure quadruple solution discussed in Ref.~\cite{Backdahl:2005uz}.}

\textcolor{black}{For the above series, if we implement the ratio test, convergence radius $R$ should depend on the ratio of the asymptotic form of the Legendre polynomials and the factorials:}
\ba
&&P_n(\cos{\theta})\sim\frac{2}{\sqrt{2\pi n \sin{\theta}}}\cos\left[ \lb n+\frac{1}{2}\rb\theta-\frac{\pi}{4}\right]\,,\\
&&\blue{n!\sim (2\pi n)^{1/2}\left(\frac{n}{e}\right)^n}\,,\\
&&\blue{n!!\sim (\pi n)^{1/2}\left(\frac{n}{e}\right)^{n/2},\quad \text{when n is even}}\,,\\
&&\blue{n!!\sim (2n)^{1/2}\left(\frac{n}{e}\right)^{n/2},\quad \text{when n is odd}}\,.
\ea
Using the above expressions, then, for example, for the first series in Eq.~\eqref{eq:USeries}, we get
\begin{align}
R&=\lim\limits_{n \to \infty} \left|\lb\frac{\mathcal M_n}{\mathcal M_{n+1}}\rb\frac{P_{6n+2}(\theta)}{P_{6n+8}(\theta)}\right|\,\nonumber\\
&=\frac{2500}{\blue{2187}}\lim\limits_{n \to \infty}\sqrt{\frac{3n+4}{3n+1}} \,\left|\frac{\cos\left[\frac{\pi}{4}-\lb\frac{5}{2}+6n\rb\theta\right]}{\cos\left[\frac{\pi}{4}-\lb\frac{17}{2}+6n\rb\theta\right]}\right|\,\nonumber\\
&=\frac{2500}{\blue{2187}}\lim\limits_{n \to \infty}\left|\frac{\cos\left[\frac{\pi}{4}-\lb\frac{5}{2}+6n\rb\theta\right]}{\cos\left[\frac{\pi}{4}-\lb\frac{17}{2}+6n\rb\theta\right]}\right|\,.
\end{align}
The limit above exists for several angles such $0,\,\pi/3,\,\pi/6$, and $\pi/2$; for each case, we find $\blue{R=2500/2187}$, which is close to unity. The same applies to the other two series in Eq.~\eqref{eq:USeries}. 
However, for other angles, $A_n$ oscillates as $n$ increases so that there is no single converged limit for such ratios. This example reflects the fact that the ratio test applies for rather limited cases where $A_n$ does not have asymptotic oscillatory behavior in $n$. On the other hand, if we use the root test for Eq.~\eqref{eq:USeries}, we find $R=2500/2187$ for any angle $0<\theta<\pi$. As a result, Eq.~\eqref{eq:USeries} converges when $\kappa^6<(2500/2187)$ which implies $\kappa < (2500/2187)^{1/6}$ or $r>(2187/2500)^{1/6}M_2^{1/3}$. Let us define,
\be
\blue{\mathcal R=(2187/2500)^{1/6}}
\ee

Note that here both ratio and root tests provide us with the magnitude of the radius of convergence. However, if we extend the metric function to the complex plane of $r$, it may not be singular for all $r$ with the same magnitude. For example, the singularity that limits the convergence behavior may locate at a negative $r$, so that the metric actually has an analytical continuation through $r=\mathcal R\, M_2^{1/3}$. In order to check whether it is the case, one needs to verify whether $U$  blows up at $r=\mathcal R\, M_2^{1/3}$ by plotting $U$ up to high order terms of $n$. If the magnitude of $U$ rises up rapidly at that location, it is an indication of singularity, which is indeed the case for Eq.~\eqref{eq:USeries} for large $M_2$.
In this particular example, the convergence radius can indeed be determined unambiguously for arbitrary $\theta$ with analytical arguments. Let us consider the first sum in Eq.~\eqref{eq:USeries}:
\be\label{eq:UFirstSeries}
\sum_{n=0}^{\infty}\mathcal M_n P_{6n+2}(\cos{\theta})\,\kappa^{6n+3}
\ee
For large $n$, each term in the sum can be expressed as
\ba
&\mathcal M_n P_{6n+2}(\cos{\theta})\,\kappa^{6n+3}\propto \frac{\blue{\mathcal R^{6n}}}{n^{3/2}\sqrt{\sin{\theta}}}\cos[(6n+\frac{5}{2})\theta-\frac{\pi}{4}]\kappa^{6n+3}\nonumber\\
&\propto \Re\left[\frac{1}{n^{3/2}}\lb e^{6i\theta}\kappa^6\blue{\mathcal R^6}\rb^n e^{i (\frac{5}{2}\theta-\frac{\pi}{4})}\right]\,,\quad 0<\theta<\pi
\ea
As a result, Eq.~\eqref{eq:UFirstSeries} becomes
\ba
&\Re &\left[ e^{i (\frac{5}{2}\theta-\frac{\pi}{4})}\sum_{n=1}^{\infty}\frac{1}{n^{3/2}}\lb e^{6i\theta}\blue{\mathcal R^6}\kappa^6\rb^n\right]+\text{Finite Part}\nonumber\\
&=&\mathrm{Li}_{\frac{3}{2}}(z)+\text{Finite Part}\,.
\ea
Where the finite part is related to the small $n$ contribution to Eq.~\eqref{eq:UFirstSeries} and $z$ is defined as $z=e^{6i\theta}\kappa^6\blue{\mathcal R^6}$. Similarly, the second and third sum in Eq.~\eqref{eq:USeries} can be expressed in terms of polylogarithm function $\mathrm{Li}_{\frac{1}{2}}(z)$. These polylogarithm functions have branch cuts along the real axis from $z=1$ to $\infty$, meaning the solution of $U$ that corresponds to $|z|>1$ is not a simple connected solution. As we extend the multipole spacetime solution towards the origin with real $r$ but different angle $\theta$, the solution can be summarised by this function that is defined in the complex $z$ plane. If the solution in the complex $z$ plane is no longer continuous because of the branch cut, the physical solution written in terms of $(r,\theta,\phi)$ is also not continuous. Therefore, the continuity requirement sets up the convergence radius as $\kappa=\blue{\mathcal R}$, which reconfirms that the convergence radius of $U$ is at $M_2^{1/3}\blue{\mathcal R}$. In general, the above analysis can be applied for cases that $a_n, A_n$ asymptote
\begin{align}
a_n, A_n \sim {\rm Re}\left ( \frac{e^{i n \alpha +\beta}}{n^\gamma R^n} \right )
\end{align}
where $\alpha, \beta, \gamma, R$ are all constants. The resulting convergence radius is $R$.

\begin{figure*}[!htb]
\includegraphics[width=\linewidth]{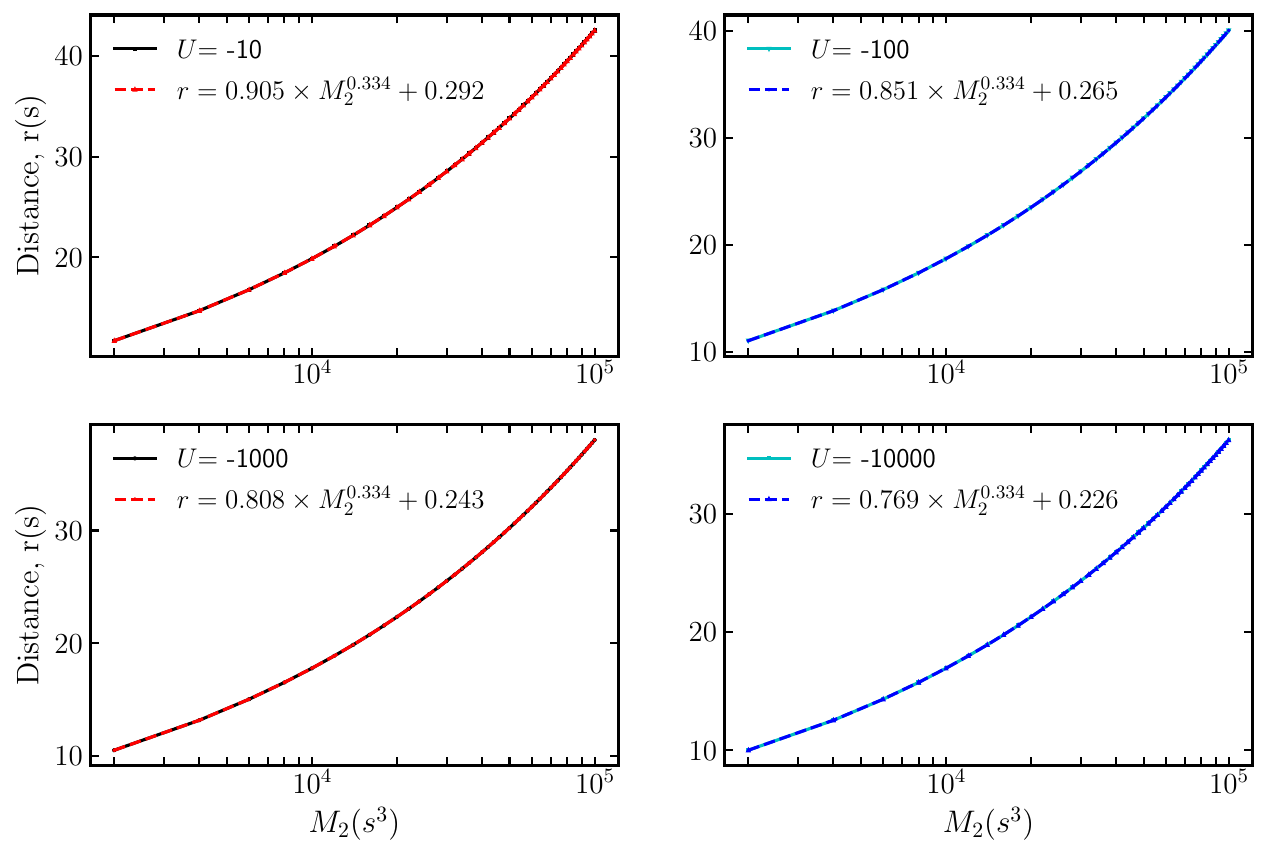}
\caption{Distance (r) from the source vs. quadrupole moment ($M_2$) plots along the symmetry axis $\theta=0$ keeping magnitudes of metric function $U$ fixed and $M_2 \in [2000,100000]\,(s^3)$. $U$ has been computed up to order $\blue{n=50}$ defined in Eq.~\eqref{eq:U}. In each plot, we have chosen a large value for $U$ and varied the quadrupole moment $M_2$ and plotted the location where U reaches such a magnitude for corresponding $M_2$. The plots follow a power law relationship of $r=a M_2^c+b$, where $a, b, c$ are constants and $c\approx 1/3$. Such a relationship further strengthens the analytical result obtained from root and ratio tests for Eq.~\eqref{eq:USeries} that location of divergence of $U$ scales as $M_n^{1/(n+1)}$ when the multipole moment $M_n$ is large.}
\label{fig:upperbound}
\end{figure*}

The convergence radius discussed so far is for Eq.~\eqref{eq:USeries}, which omits terms with lower powers of $M_2$ and is not an exact solution of Einstein equations. To see if the above convergence radius is consistent with the actual solution of $U$ in Eq.~\eqref{eq:U}, we consider the upper bound test. For various upper bounds of $U$, we vary $M_2$ and find the location $r$ with the corresponding upper bound. The location of the upper bound follows a power law dependence on $M_2$ with $r=a M_2^n+b$ where a and b are constants and $n\approx 1/3$. This is demonstrated in Fig.~\ref{fig:upperbound}, with $M_2$ vs. r plot with various upper bounds of $U$ showing the same exponent $n$.

The scaling law of $M_2^{1/3}$ is interesting as it is generally smaller than the expectation from Newtonian gravity. Indeed for a  source with mass $M$ and size $L$, one would expect that the largest $M_2$  is achieved by placing the mass on the boundaries, which leads to $M_2 \sim M L^2$. This means the Newtonian intuition is not necessarily correct in the case of a GR definition of multipole moments and their relation with the size of an object, suggesting such a difference with Newtonian physics is possibly due to strong-field effects. In particular, the ``ultra-compact" solution reaching the lower bound in size may be highly nontrivial, which may not have a post-Newtonian, fluid-type source construction. More studies are required to find an explicit  construction of the source that saturates the bound. Similarly, observing a few higher-order moments, we find that if the metric is dominated by the multipole moment $M_n$, the convergence radius scales as $M_n^{1/(n+1)}$.

\begin{figure}[!h]
\includegraphics[width=8.5cm]{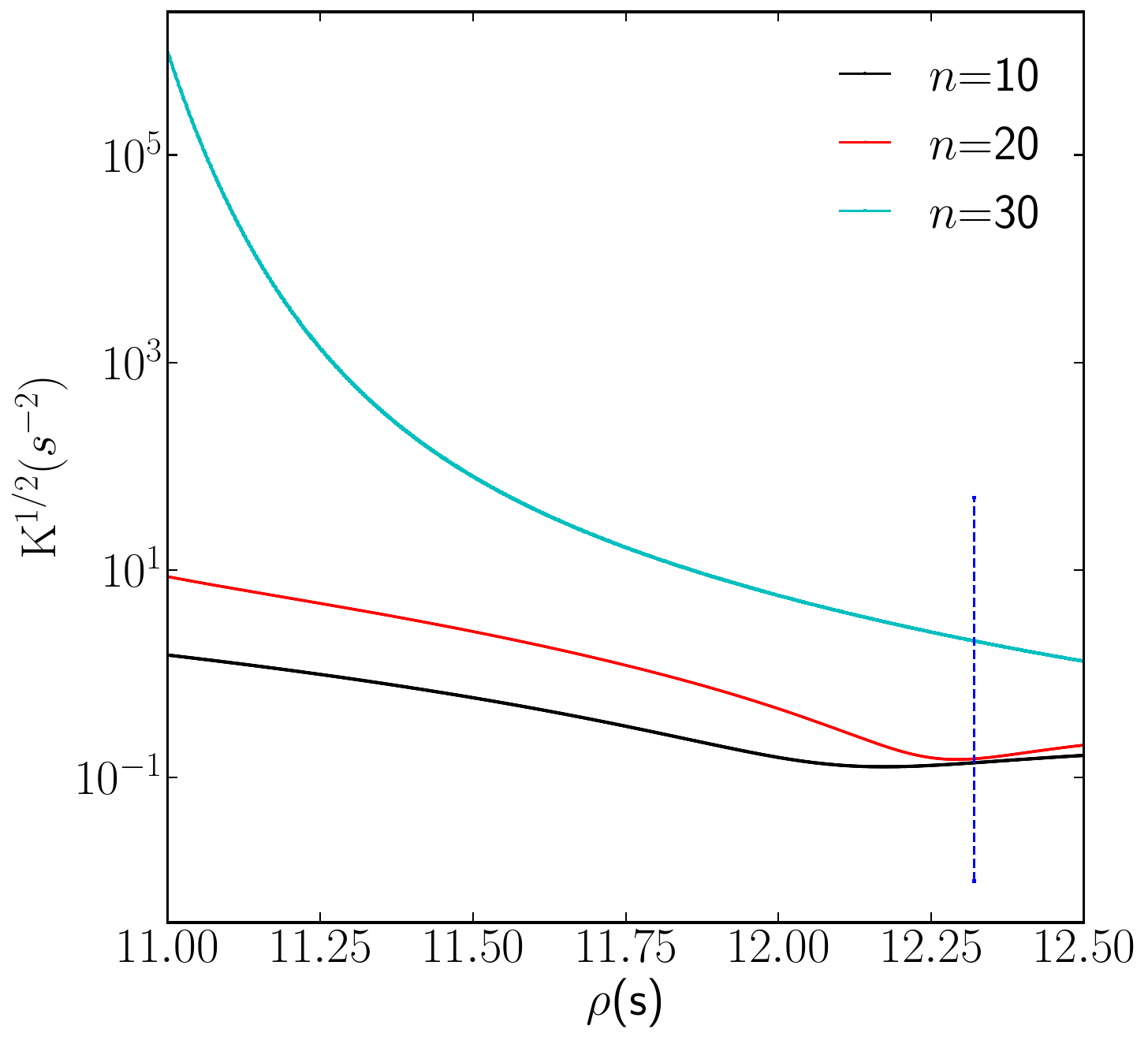}
\caption{Square root of Krestchmann curvature ($\sqrt{K}$) vs. distance ($\rho$) plot on equatorial plane $z=0$ in Weyl-Papapetrou coordinates for the metric described in Eqs.~\eqref{eq:U}-\eqref{eq:gamma}. We have chosen $M=1$ and $M_2=2000$ and ignored other multipole moments. The vertical dashed line shows the location $\blue{\mathcal R M_2^{1/3}}$ on the horizontal axis. Curvature increases rapidly when $\rho$ is smaller than the convergence radius and increases with order $n$ in Eqs.~\eqref{eq:U}-\eqref{eq:gamma}.}
\label{fig:ks}
\end{figure}

\begin{figure}[!h]
\includegraphics[width=8.5cm]{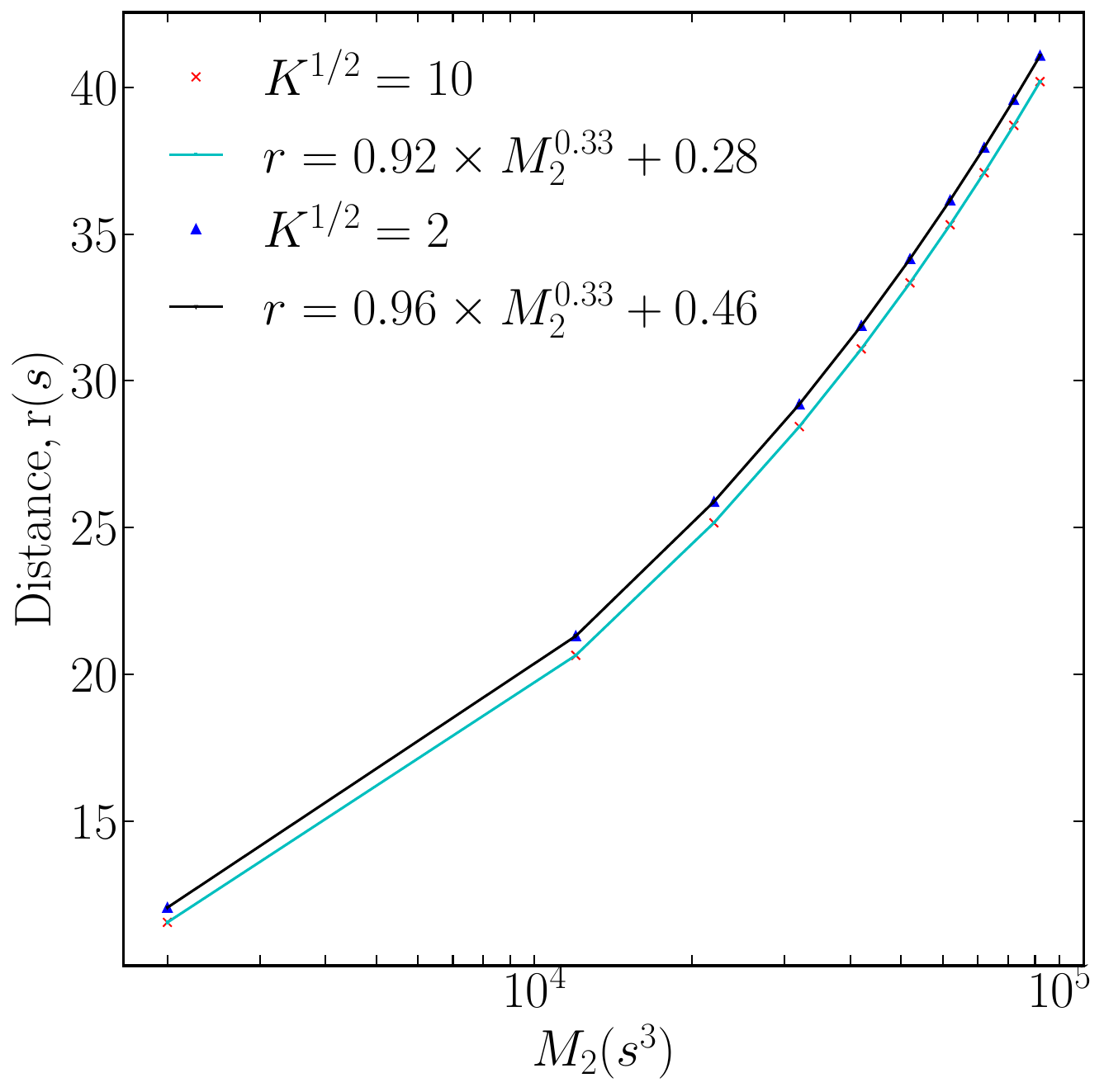}
\caption{\blue{Location ($r$) of curvature threshold vs. $M_2$ plot on the $z=0$ plane for the metric in Eq.~\eqref{eq:met_static} dominated by the quadrupolar moment. We perform upper bound tests with the square root of Kretschman curvature $\sqrt{K}=2$ and $\sqrt{K}=10$. We find that similar to the upper bound test with metric function $U$, the location of the curvature threshold follows the power law $r=a M_2^c+b$, with the exponent $c$ being close to $1/3$.}}
\label{fig:ks2}
\end{figure}

While the above convergence tests provide us with information on the location of divergence of the metric components, metric components are coordinate-dependent quantities, so that the divergence could be the result of a choice of coordinates. As a result, we need to check the behavior of curvature invariants at the convergence radius to see if they are indeed divergent. If a curvature invariant blows up at the convergence radius, it is an indication that the vacuum solution breaks down and that the convergence radius indeed limits  the size of the source. For this purpose, we compute the Krestchmann scalar $K=R^{abcd}R_{abcd}$ for the metric in Eqs.~\eqref{eq:U}-\eqref{eq:gamma} in the limit of $M_2\gg M^3$. The behavior of the curvature, of course, depends on the number of terms we keep in $U$ and $\gamma$. While for consecutive orders, the magnitude of $K$ may display an increasing or decreasing trend at the location of convergence radius, as we keep higher and higher order terms in $n$, $K$ unambiguously increases significantly. This suggests that the curvature invariant will also diverge at the convergence radius of $U$ if we keep a sufficient number of terms in the expansion. This feature is shown in Fig.~\ref{fig:ks} by comparing $\sqrt{K}$ considering metric for $n$ up to $n=10$, \blue{$n=20$}, and \blue{$n=30$}. \blue{Furthermore, performing upper-bound tests with $\sqrt{K}$ also shows that the convergence radius scales as $M_n^{1/(n+1)}$. In Fig.~\ref{fig:ks2}, we show the upper bound tests with curvature $\sqrt{K}=2$ and $\sqrt{K}=10$ for $M_2\gg M^3$ case. For both upper bounds, the location of the curvature scales as $M_2^{1/3}$.}

\subsection{Kerr Mimickers}\label{sec:3B}
We now consider the limit that spacetime multipoles are weakly perturbed from Kerr values. This scenario is particularly useful for the tests of black hole mimickers. For simplicity, we assume  the spacetime only deviates from Kerr in its quadrupole moment. In other words, $M_n=M (ia)^n$ for all $n$ except $M_2=-M a^2+\delta M_2$, where we consider $\delta M_2$ to be small in magnitude compared to the Kerr quadrupole moment $-Ma^2$. For numerical computations and presentations in the plots, we use the normalization that $M=1$. \blue{Following the procedure discussed in the last paragraph of Sec.~\ref{sec:2b}, we compute $m_n$ in terms of moments up to $n=25$ and then compute the metric functions.} The power-law expansion of the {non-Kerr} part of the metric is accurate up to $\mathcal O (1/r^{26})$ in terms of a quasi-Cartesian set of coordinates, while we take the background to be exact Kerr. Here $a$ is still defined as $J/M$, with $J$ being the angular momentum of the source. We will explore how the radius of convergence depends on $\delta M_2$ and the spin parameter $a$. In almost all the analyses here and the rest of the paper, we consider $\delta M_2$ to be positive; as a result, we will refer to the condition of small Kerr quadrupole modification as $\delta M_2 < M a^2$.

Because the metric is only weakly perturbed from Kerr, we find it convenient to convert the metric expressed in Weyl-Pappapetrou coordinates to Boyer-Lindquist-like coordinates to compute the convergence radius. Notice that the coordinate transformation is the same as the one that takes a Kerr metric from Weyl-Papapetrou coordinates to  Boyer-Lindquist coordinates:
\be\label{eq:WptoBL}
\rho=(r_{BL}^2-2M r_{BL}+a^2)^{1/2}\sin{\theta_{BL}}\,,\quad z=(r_{BL}-M)\cos{\theta_{BL}}\,.
\ee
We also transform $\varphi$ as $\varphi\rightarrow -\varphi$. Note that the above coordinate transformation works only outside the Kerr horizon so that $\rho$ is real and positive.

\begin{figure}[!htb]
\includegraphics[width=8.5cm]{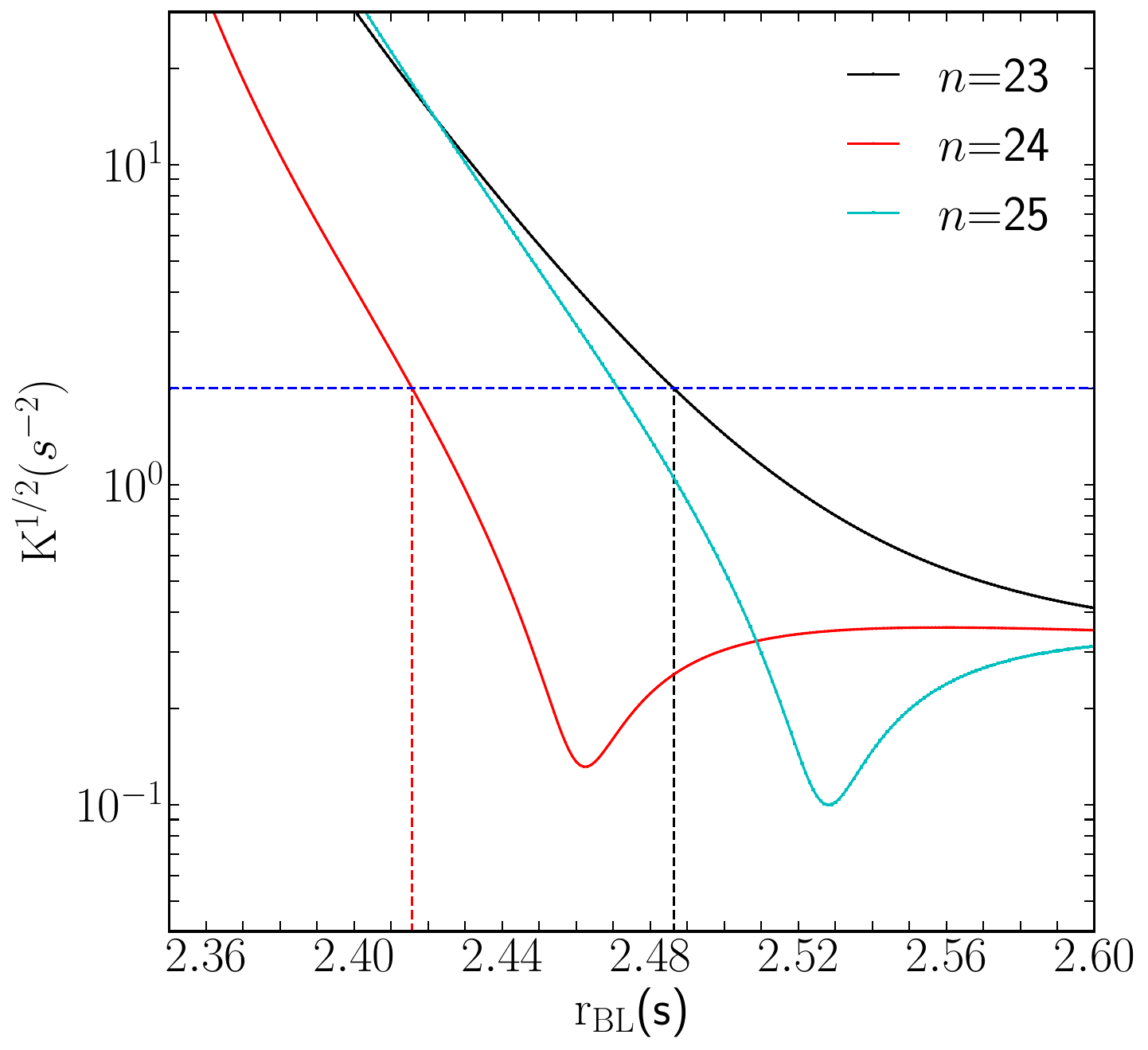}
\caption{Square root of Krestchmann curvature ($K^{1/2}$) vs distance ($r_{BL}$) plot in Boyer-Lindquist-like coordinates on the equatorial plane ($\theta_{BL}=\pi/2$) at various orders in inverse radial distance. The magnitude of curvature at a location increases with order in general but may fluctuate for consecutive orders. The horizontal range between the black dotted line and the red dotted line shows the variation of the location of $\sqrt{K}=2$ for orders $n=23,\,24,\,25$. Spin is chosen as $a=0.3$ and non-Kerr modification $\delta M_2=0.08$. Note that by order $n$, we mean the highest order of multipole moments included in the metric computation is of order $n$.}
\label{fig:modkerrcurv}
\end{figure}

\begin{figure}[!htb]
\includegraphics[width=8.5cm]{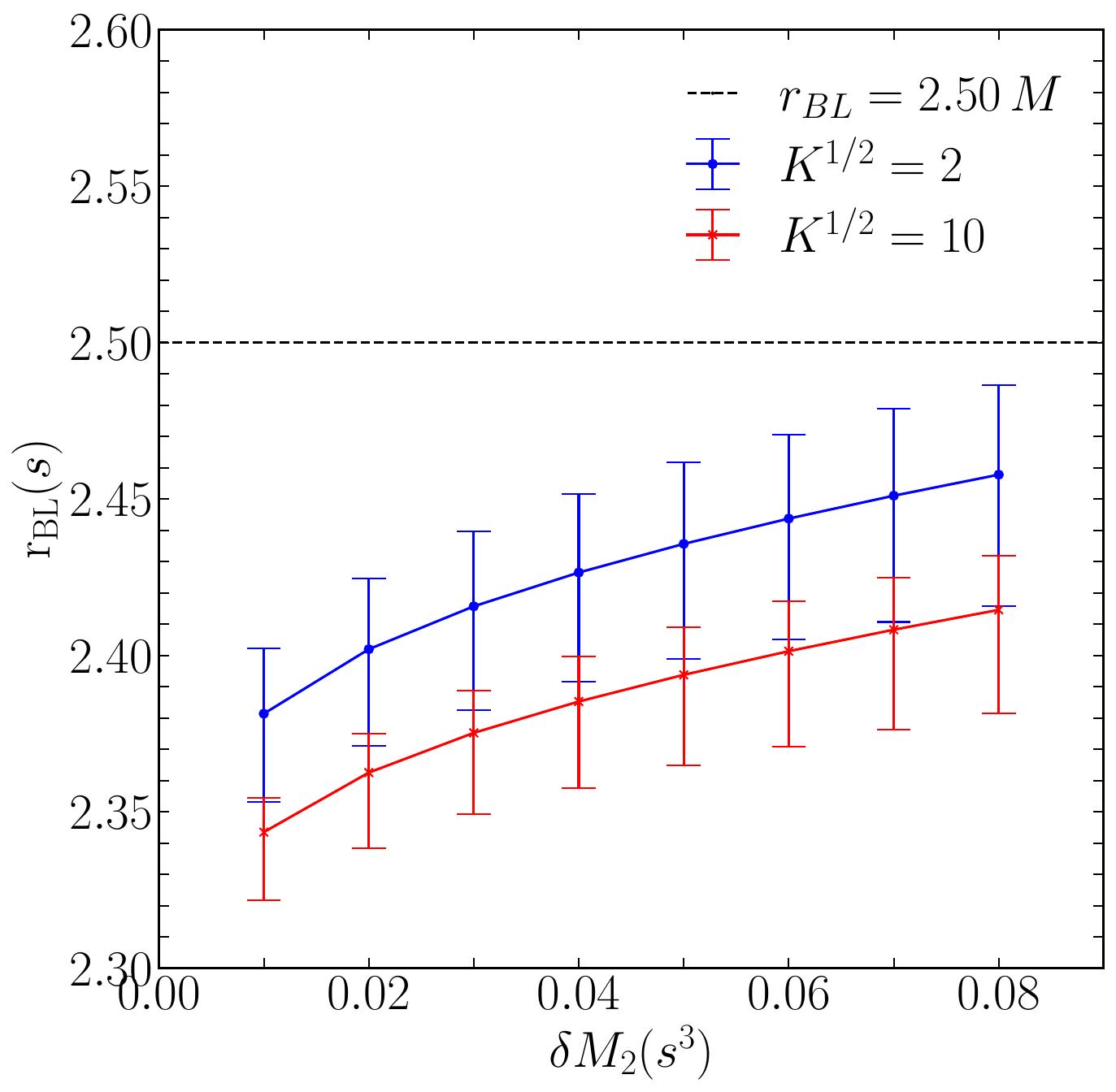}
\caption{Location ($r_{BL}$) of curvature threshold vs. $\delta M_2$ plot on the equatorial plane for spin $a=0.3$. The red line denotes the location of $\sqrt{K}=10$ and the blue line denotes the location of $\sqrt{K}=2$ for various non-Kerr modifications. Each point shows the mean value of location obtained from orders $n=23,\,24$ and $25$. The vertical ``error bar" corresponding to each point shows how much the location varies depending on orders. The horizontal black dashed line corresponds to \blue{$r_{BL}=2.50 M$} which is the cutoff radius considered for generating BH shadows in Sec.~\ref{sec:4}. For a comparison, at \blue{$r_{BL}=2.50 M$}, from Fig.~\ref{fig:modkerrcurv}, we obtain \blue{$\sqrt{K}\approx 0.8$} by averaging over the three orders, while for Kerr solution, $\sqrt{K}=0.4$ at the equatorial prograde photon orbit radius.}
\label{fig:rBLvsdeltam2_curv}
\end{figure}

In the original discussion of the No-Hair Theorem of Kerr black holes, it was realized that the homogeneous perturbation of the spacetime that introduces a nonzero modification to multipole moments at infinity has to blow up at the black hole horizon (a curvature singularity) \cite{Israel:1967wq}. This fact is also used regarding the calculation of black hole Tidal Love numbers \cite{Binnington:2009bb}. In the small $\delta M_2$ limit, the perturbed spacetime we compute here is essentially the ``blowing-up-at-horizon" piece of the homogeneous solution. Of course, as $\delta M_2$ increases, the radius of singularity deviates from the black hole horizon. This is consistent with the curvature of our metric presented in Fig~\ref{fig:modkerrcurv}, for which we consider an upper-bound test in Fig.~\ref{fig:rBLvsdeltam2_curv}. Ideally, one would want to study the curvature in the limit of very large $n$. In our case, the highest three orders we used for computing curvatures are shown in Fig~\ref{fig:rBLvsdeltam2_curv}. For a certain upper bound on $\sqrt{K}$, the location where we reach the upper bound fluctuates with orders. Setting an upper bound and computing the distance from the source to the location of the bound for each of the three orders, we compute a range of fluctuation of the curvature.

Qualitatively, the distance from the source to the location of curvature bound (see Fig.~\ref{fig:rBLvsdeltam2_curv}) increases with increasing magnitude of $\delta M_2$. This is also true for negative non-Kerr deviations, and choosing a different curvature upper bound produces a similar result. Furthermore, for any small $\delta M_2$, a large curvature threshold can be achieved outside the Kerr horizon, while for Kerr such magnitudes of curvature are located inside the horizon. If a curvature singularity exists near the source,  the location of singularity should constrain the minimal size of the source. Although the radius values presented in  Fig.~\ref{fig:rBLvsdeltam2_curv} are coordinate dependent, the monotonic trend still suggests that bigger objects create larger $\delta M_2$ for a fixed spin.
\section{ Shadow Measurement}\label{sec:4}
One way to search for deviations from the Kerr metric in spacetimes with different sets of multipole moments is to measure the shadows (or critical curves) in such spacetimes. Searching for the relation between the critical curve and the multipole moments has recently been discussed in \cite{Glampedakis:2023eek}, which  is mostly focusing on photon orbits on the equatorial plane for a set of spacetimes with slow spins or linearized metric perturbations. In this work, we apply the explicit metric constructed in Sec.~\ref{sec:3B} and numerically compute the black hole shadows/critical curves. The technical term ``\emph{shadow}" represents the projection of the critical photon sphere of the central object on the camera plane being observed by a distant observer. If the central object has a hard surface, the emission from accreted material on the surface likely significantly changes the radio image \cite{Broderick:2005xa}. If the size of the surface is larger than the radius of the critical curve of the spacetime, there are likely no ``photon rings" showing up in the radio image, similar to the ones predicted for black holes.

In this section, we compute the shadow of the central object with the stationary Weyl-Papapetrou metric discussed in Sec.~\ref{sec:3B}. Using the coordinate transformations in Eqs.~\eqref{eq:WptoBL}, we transform the metric to the Boyer-Lindquist-like coordinates. As a sample problem, we choose a spin of $a=0.3$ setting $M=1$ and consider two different values of $\delta M_2$ of 0.01 and 0.08. We choose such a set of parameters so that the photon ring of Kerr spacetime with the corresponding spin value $a=0.3$ (prograde photon ring radius is $r^-_{ph} = 2.63$ \footnote{Note that the unstable spherical Kerr photon orbits have radius $r^{-}_{ph}\leq r \leq r^{+}_{ph}$ where $r_{ph}^-$ and $r_{ph}^+$ are the circular prograde and retrograde photon orbits in the equatorial plane. We can find these two radii analytically using $r_{ph}^{\pm}=2M [1+\cos{[(2/3)\cos^{-1}(\mp a/M)]}]$~\cite{Teo2003}.\label{kerrphoton}}) is outside the curvature bound in Fig.~\ref{fig:rBLvsdeltam2_curv}. Note that the actual photon ring of our metric is not necessarily that of Kerr and it is not guaranteed that there will be a photon ring outside the source for this set of parameters. However, later in this section, we show how we find a cut-off radius in our shadow simulation compatible with this assumption and also with the qualitative measure of the upper bound for the curvature Fig.~\ref{fig:rBLvsdeltam2_curv}.

In general, we do not expect a  Carter-like constant for generic multipole spacetimes. As a result, the geodesic equations are not necessarily separable to be solved analytically in a straightforward way. To calculate the shadow of our case study spacetime we solve the null geodesics backward in time fully numerically \footnote{ Numerical integrations are performed for all eight equations corresponding to the four Boyer-Lindquist coordinates ${t, r, \theta, \phi}$ and their time derivatives, without resorting to any constants of motion such as energy or angular momentum.}. The method is commonly known in the field also as ``backward null ray tracing". For this purpose, we modify the ray-tracing part of the publicly available code, ``Odyssey" \cite{Pu:2016eml} (excluding the radiative transfer part).
In addition, since the metric components are expressed in series expansions in the inverse radial coordinate involving numerous terms, we interpolate the metric data and then compute the Christoffel symbols numerically, e.g.  for the right-hand-side of the geodesics equation:
\begin{equation}
    \frac{d^2 x^{ \ \mu}}{d \lambda^2} = \Gamma^{ \ \mu}_{\alpha \ \beta} \ \frac{d x^{ \ \alpha}}{d \lambda} \frac{dx ^{ \ \beta}}{d \lambda} \equiv F^{ \ \mu }_{int} (r,\theta)
\end{equation}

In the last step, we numerically integrate the geodesic equations (the left-hand side of the equation above) in the relevant domain of interest. The interpolant functions for the metric components are accurate up to $10^{-6}$ in the fractional difference. It is also worth mentioning that this makes our code capable of calculating the shadow for any arbitrary axisymmetric spacetime. It is also easily extendable to more general spacetimes.  
 In the following, we will discuss the initial conditions of the rays and the choice of integration domain in more technical detail.

\subsection{Numerical Shadow Details}\label{sec:4A}
To calculate the shadow of the central object numerically, we employ the backward ray tracing method, a well-established technique in which the null geodesic equations are numerically integrated in reverse time. First, we establish a correspondence between the initial conditions of each ray on the camera in the observer frame (or the camera frame) and the central object's frame (the object responsible for creating the shadow). We refer to points on the camera plane as ``pixels". The coordinates tied to the central object are the Boyer-Lindquist-like coordinates ($r_{BL},\,\theta_{BL},\,\phi_{BL}$). However, in order to keep notations simple we will denote them by ($r,\,\theta,\,\phi$) without the subscript ``BL".

The transformation procedure that relates a pixel in the camera frame to that of the central object's frame follows the methodologies outlined in the references~\cite{Younsi_2016, Pu:2016eml} as briefly described below:
\begin{eqnarray}
x_{\mathrm{CO}} &=& \sqrt{r_O^2 + a^2} \sin \theta_O - y_{c} \cos \theta_O  \,,\label{x0} \\
y_{\mathrm{CO}} &=& x_{c} \,, \label{theta0} \\
z_{\mathrm{CO}} &=& r_O \cos \theta_O \,. \label{phi0}
\end{eqnarray}
The subscripts ``O" and ``CO" refer to ``Observer" and ``Central Object", respectively. Specifically, ($r_O,\,\theta_O,\,\phi_O$) denotes the location of the observer in the central object's frame. The coordinates $(x_{\mathrm{CO}}, \,y_{\mathrm{CO}},\,z_{\mathrm{CO}})$  are Cartesian coordinates of the pixels in the central object's frame and the coordinates $(x_c, y_c, z_c)$ are the pixel coordinates in the camera frame, with $z_c=0$. In addition,  the $z$ coordinate in both the observer's and central object's frame is chosen to be aligned. For details of the coordinate transformation between these two frames readers can refer to \cite{Pu:2016eml}. Since the spacetime in our case study still has axial symmetry, we can set $\phi_O = 0$ without any loss of generality. We then transform $(x_{\mathrm{CO}}, \,y_{\mathrm{CO}},\,z_{\mathrm{CO}})$ to Boyer-Lindquist-like coordinates in the central object's frame, which let us relate any sets of pixel coordinates (as our initial positions of the rays) to the BL coordinates in the CO frame:
\begin{eqnarray}
\ \! r &=& \sqrt{ \frac{u +\sqrt{u^{2}+4 a^{2} z_{\mathrm{\mathrm{CO}}}^{2}}}{2} \,, \label{r0}} \\
\theta &=& \cos^{-1} (\frac{z_{\mathrm{CO}}}{r}) \,, \label{theta0} \\
 \phi &=& \tan^{-1}(\frac{y_{\mathrm{CO}}}{x_{\mathrm{CO}}}) \,, \label{phi0} \\
t &=& 0 \,,
\end{eqnarray}
where $ u = x_{\mathrm{CO}}^2 + y_{\mathrm{CO}}^2  + z_{\mathrm{CO}}^2 - a^2$. Accordingly, the initial conditions for velocities  are:
\begin{eqnarray}
\ \dot{r} &=& -\frac{r\,\mathcal{R} \sin\theta \sin\theta_{\mathrm{O}} \cos \phi+\mathcal{R}^{2} \cos\theta \cos\theta_{\mathrm{O}}}{ (r^{2} +a^{2}\cos^{2}\theta)  } \,, \label{rdot0} \nonumber\\ \\
 \ \dot{\theta} &=& -\frac{\mathcal{R} \cos\theta\sin\theta_{\mathrm{O}} \cos \phi- r\sin\theta\cos \theta_{\mathrm{O}}}{ (r^{2}+a^{2}\cos^{2}\theta) } \,, \label{thetadot0}\nonumber\\ \\
\ \dot{\phi}&=& \frac{\sin\theta_{\mathrm{O}} \sin \phi\ \! \csc \theta}{ \mathcal{R} } \,, \label{phidot0}
\end{eqnarray}
in which $ \mathcal{R}\equiv \sqrt{r^{2}+a^{2}}$. The overhead dot denotes the derivative with respect to an affine parameter. The initial value for $\dot{t}$ is then calculated numerically for each pixel using the null condition $p^{\mu}p_{\mu} = 0$, where $p^{\mu}$ denotes the four-momentum.
\begin{figure}[htb]
  \centering
  \subfigure{
    \includegraphics[width = 8.0cm]{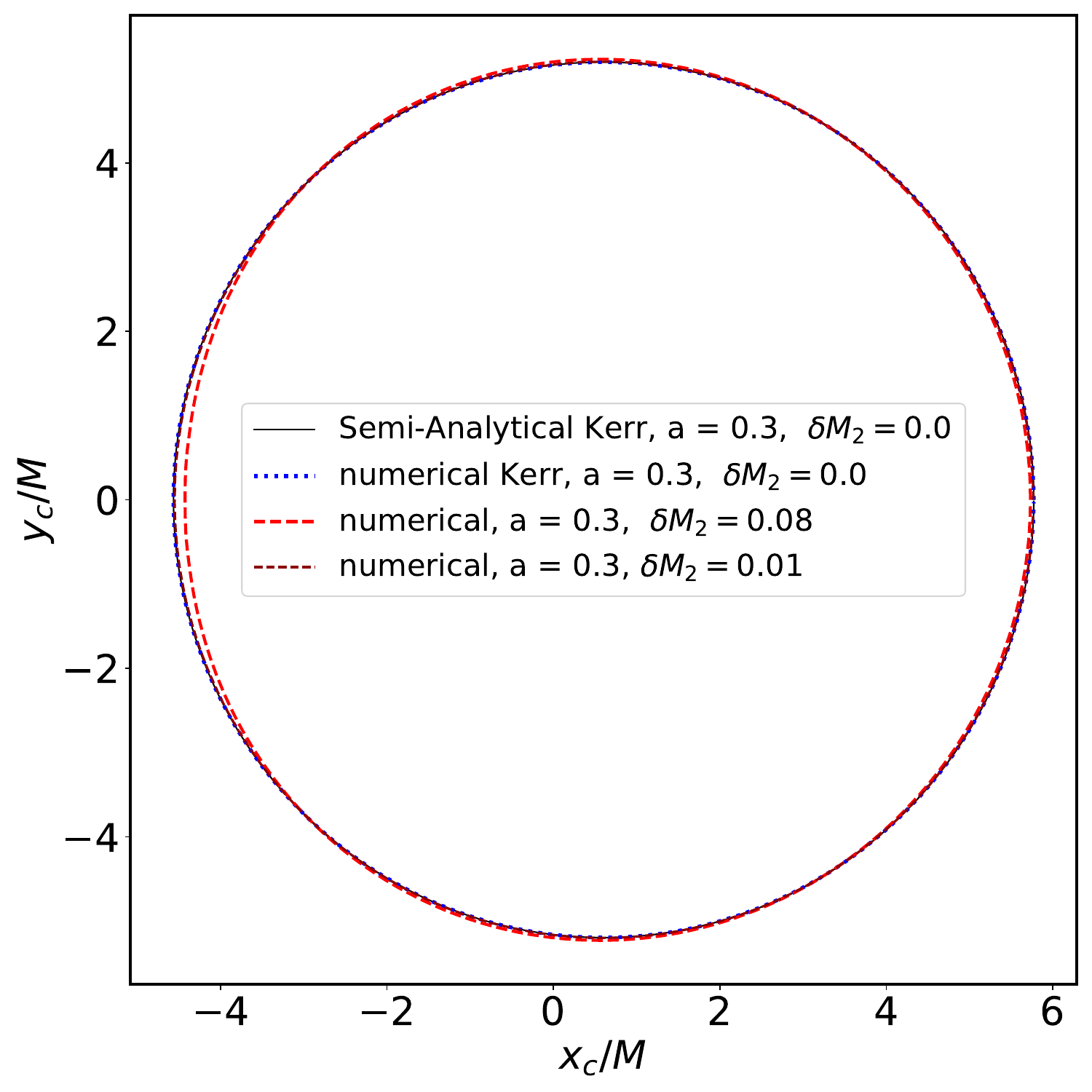}
  }
  \subfigure{
    \includegraphics[width = 8.0cm]{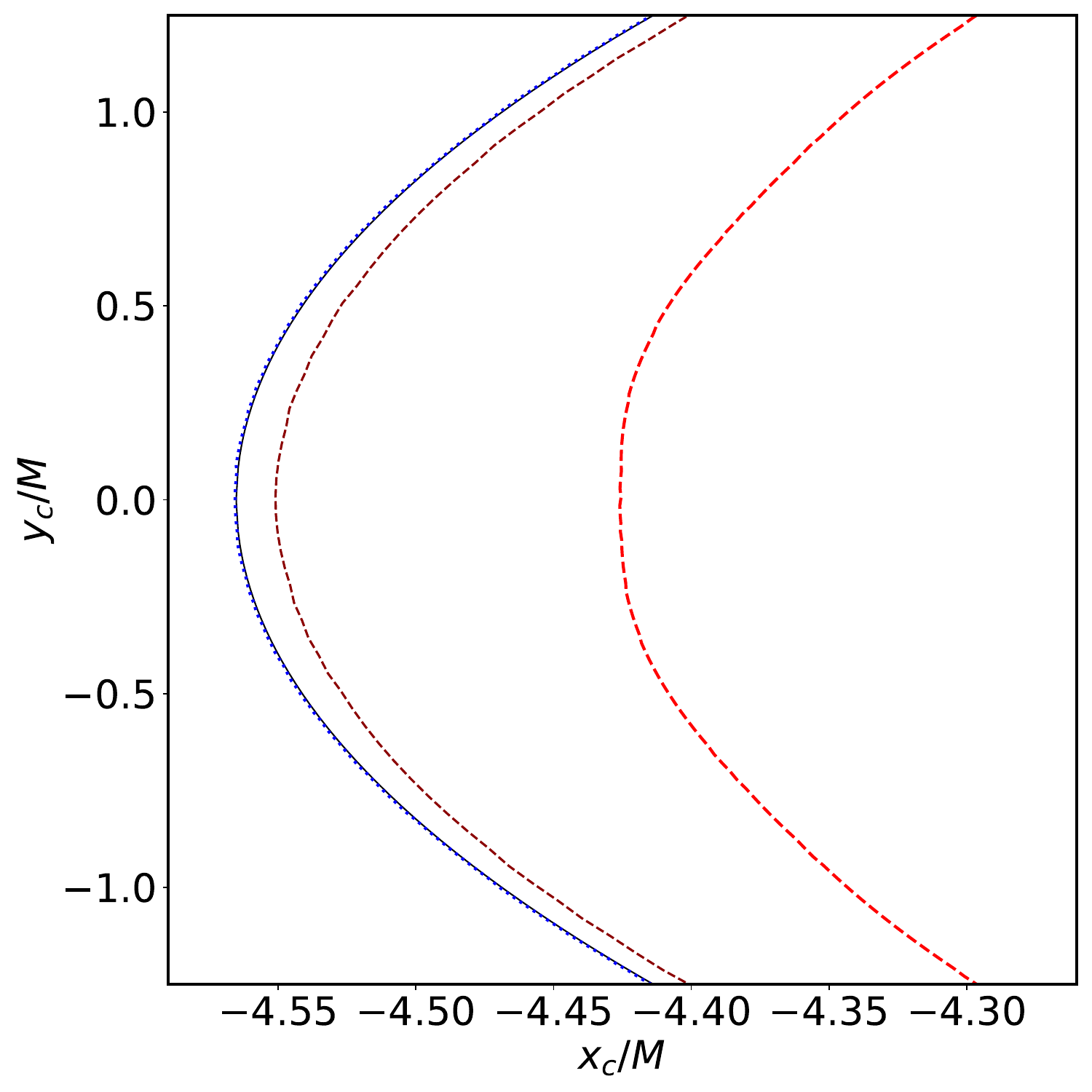}
  }
  \caption{Critical photon ring seen by an observer in the camera frame on the equatorial plane ($\theta_O = \pi/2$). The top panel shows the shadow edge of a Kerr BH with $M=1$ and $a=0.3$ computed in a semi-analytical method and also using our numerical method, along with two cases of  $\delta M_2 = 0.01$ and $\delta M_2 = 0.08$. The lower panel is the zoom box for the left part of the shadow which corresponds to the co-rotating null geodesics. The Kerr shadows computed semi-analytically and numerically are almost indistinguishable, which verifies the accuracy of our numerical simulations.}
\label{fig:shadows}
\end{figure}

As mentioned previously, we compute the photon sphere of a compact object with our spacetime metric numerically using the backward ray-tracing method. Therefore, a major issue in simulating the shadow of this spacetime is the choice of the integration domain. Our integration begins at points located on the equatorial plane, with initial conditions ($\theta_0 = \pi/2$, and arbitrary $\phi_0$).
The starting radial coordinate is $r_0 = 600 M$, and integration proceeds backward to a point we designate as the cut-off radius $r_{cut}$. 
Unlike the case of a black hole, we cannot integrate all the way back to the horizon because the convergence radius in our model exceeds that of the Kerr horizon. Additionally, our metric loses its smoothness as it approaches small distances from the center of CO (due to the limited number of terms in the expansion series).

Therefore, we set a cut-off radius for our ray-tracing procedure to ensure that, upon passing $r_{\text{cut}}$, the photon would ultimately be captured by the central object (CO). To validate this choice, we experiment with different values for the cut-off radius, all of which are less than or equal to $2.63 M$ (the radius of Kerr's prograde photon ring for $a=0.3$). Specifically, we test the range \textcolor{black}{$r_{\text{cut}} = [2.63, 2.38]$} and observe that for \textcolor{black}{$r_{\text{cut}} \leq 2.5$}, the border of the shadow converges to a fixed point. As a result, we select $r_{\text{cut}} < 2.48 M$ as our integration cut-off radius in addition to the condition $\dot{r} < 0$ which ensures that the rays passing the cut-off radius would have a negative radial velocity. These conditions confirm null rays' eventual capture by the CO.
Keep in mind that this particular concern would arise only on the left side of the shadow along the equatorial line. At this location, null rays would be co-rotating with the CO's spin, and the photon sphere radius reaches its minimum value. This allows the rays to come as close as possible to the CO. While we don't have an analytical expression for the exact location of the photon sphere in our metric, its symmetry closely resembles that of the Kerr metric. Therefore, we anticipate that, even in our metric, the photon sphere radius will be larger at any point other than the left side of the shadow. For this reason, we compare our cut-off radius with the equatorial photon ring radius instead of the photon sphere.
Besides, the fact that the equatorial prograde edge of the shadow (left side of the shadow) saturates to a fixed radius, means that the surface of the CO lies within the $r_{cut}$, which can be a numerical proof of what is shown in Fig. \ref{fig:rBLvsdeltam2_curv}.

\begin{figure}[t]
    \centering
    \subfigure{
    \includegraphics[width =8.5cm]{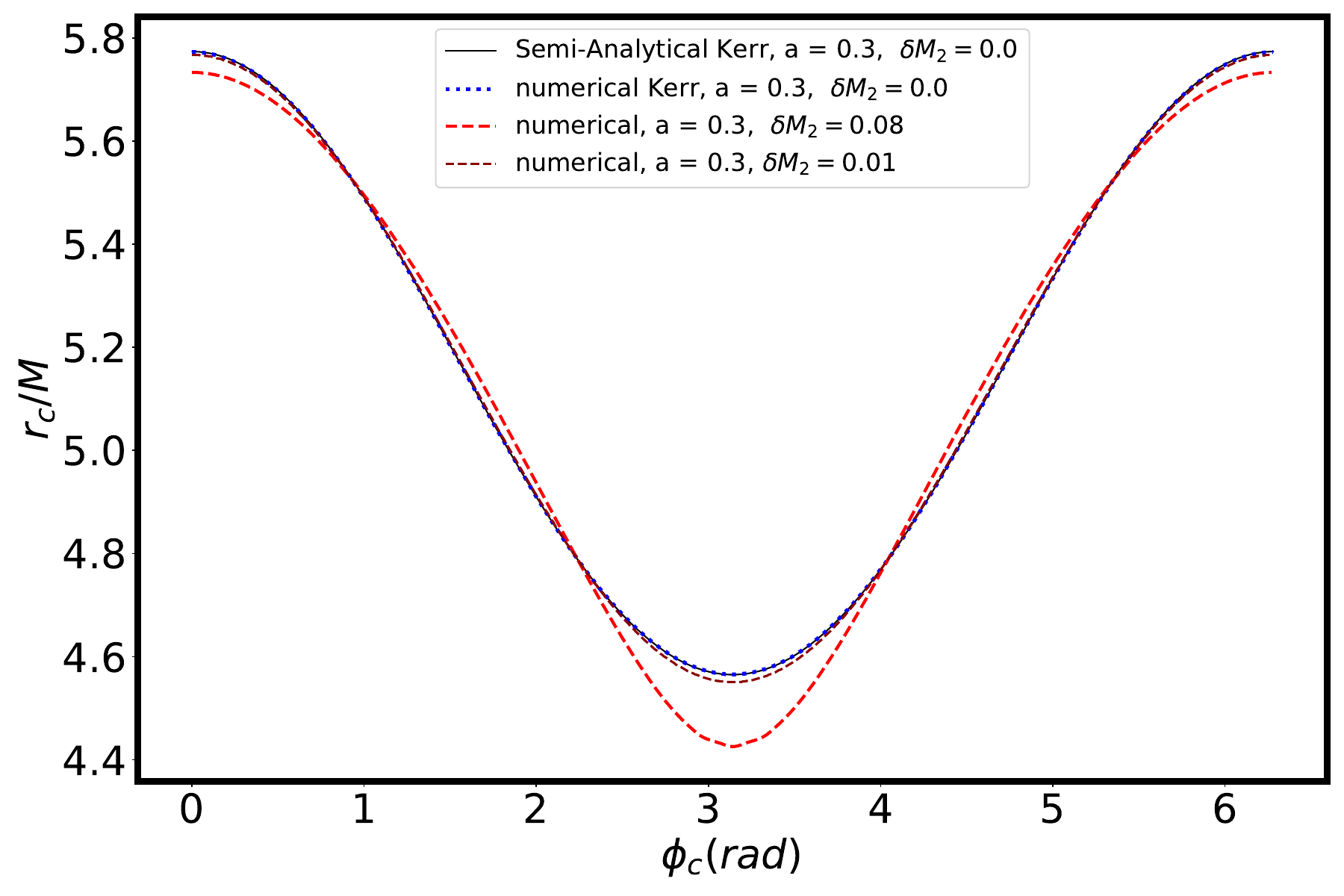}
    }
\vfill
\subfigure{
 \includegraphics[width = 8.5cm]{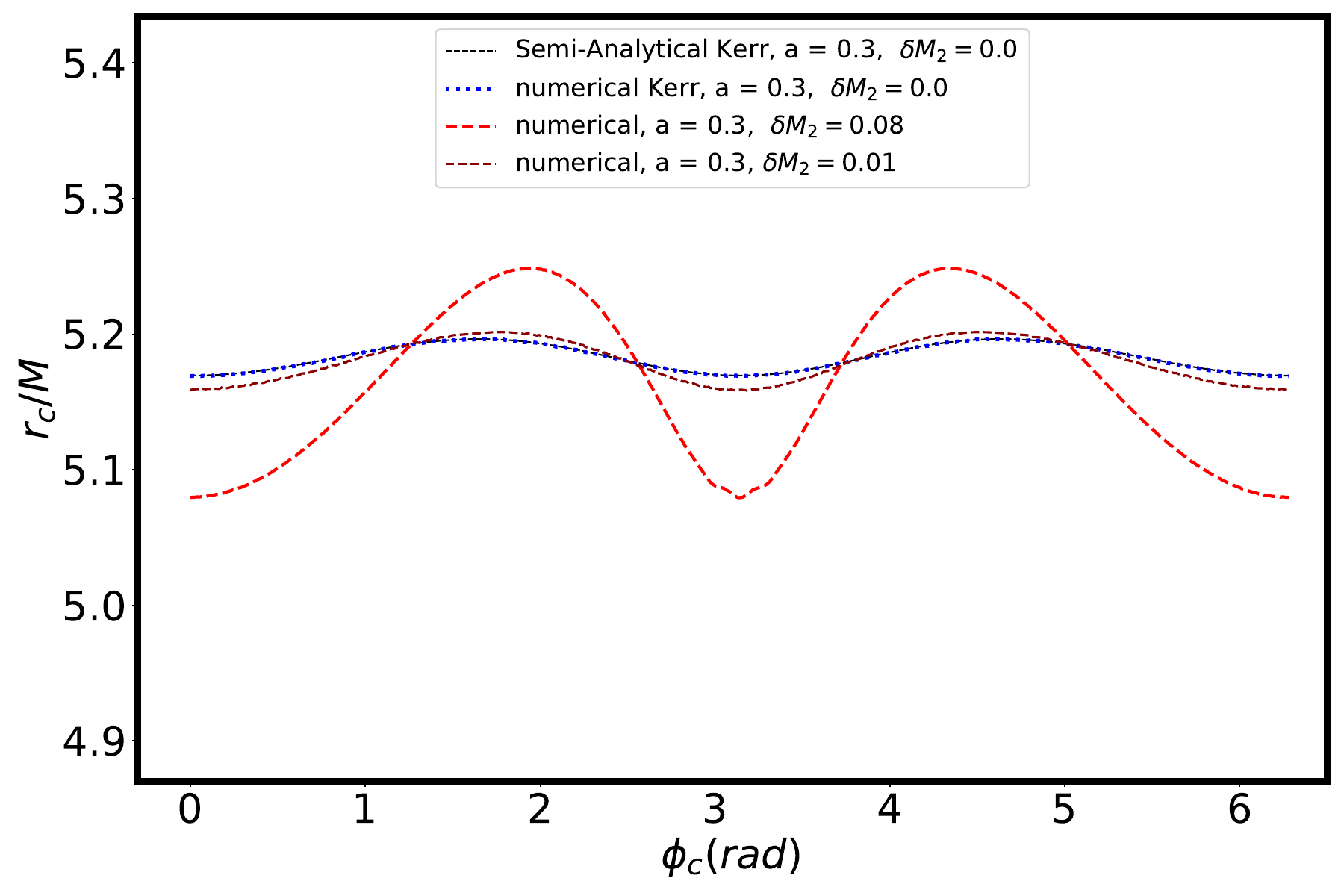}
 }
\caption{The shadow border's radius ($r_c$) as a function of the polar angle ($\phi_c$) at the center of the camera frame, where $r_c=\lb x_c^2+y_c^2\rb^{1/2}$ and $\phi_c=\arctan{(y_c/x_c)}$. In the lower panel, the shadows have been co-centered to the origin of the coordinate system in the camera frame. The small oscillation appearing in the critical curve at $\phi_c = \pi$ is due to the limited number of terms in the metric series expansion.}
\label{fig:shadows_rth}
\end{figure}
Let us now discuss the characteristics of the simulated shadow of the central object. We assume that the observer is located on the compact object's equatorial plane. For an initial quantitative assessment of the shadow boundaries obtained from our simulation, we consider the equatorial diameter of the shadow, denoted as $d_e = r_{ph}^+ + r_{ph}^-$. Here $r_{ph}^{\pm}$ represent the radii of unstable equatorial circular null rays that co-rotate ($-$) or counter-rotate ($+$) with the direction of the black hole's spin. 
Two key points warrant further discussion. First,
the left side of the shadow is shaped by rays that co-rotate with the central object (CO) in the equatorial plane, thereby approaching it more closely\footnote{The null prograde rays form an unstable circular orbit with a smaller radius than the retrograde ones.}. As these rays draw closer to the CO, deviations from the Kerr metric become increasingly evident. Consequently, we anticipate the greatest difference in the shadow to manifest along the left side of its border. Second, for comparative purposes, we have a readily available analytical expression for the radius of unstable circular orbits in the equatorial plane in the Kerr scenario, as outlined in~\ref{kerrphoton}. We find that the deviation from a corresponding Kerr spacetime shadow for the case with $\delta M_2 = 0.01$ is not noticeable \textcolor{black}{($\delta d_e = 0.2\%$)} (see Fig.~\ref{fig:shadows}), while the $\delta M_2=0.08$ case differs from the Kerr by \textcolor{black}{$\delta d_e = 1.74\%$}. Notice that the deviation in the equatorial diameter for our numerical Kerr shadow from the analytical Kerr value of that is only $\delta d_e = 6\times 10^{-5}$.

To have a better illustrative comparison between the deviations of the borders we also plot the shadow border in the polar coordinates in camera frame $(r_c, \phi)$ Fig. \ref{fig:shadows_rth}, where:
\begin{equation}
\begin{split}
r_c &= \sqrt{x_c^2 + y_c^2} \\
\phi &= \arctan (\frac{y_c}{x_c})
\end{split}
\end{equation}

A distinguishing feature of this spacetime is the deviation in the vertical size of the shadow on the camera plane---the shadow radius when $\phi_c=\pi/2$ or $\phi_c=3\pi/2$---from that of the Kerr black hole. In the case of the Kerr black hole, this vertical size remains constant and is identical to the Schwarzschild black hole's, regardless of its spin value. To show this specific characteristic, we plot $r_c$ against $\phi_c$ for $\delta M_2=0.08$, $\delta M_2=0.01$, and the corresponding Kerr $\delta M_2=0$ (see Fig. \ref{fig:shadows_rth}). We find that after co-centering the origin of the shadows on the camera plane (lower panel of Fig. \ref{fig:shadows_rth}), $\delta M_2=0.08$ produces a larger deviation in the vertical size of the shadow from that of Kerr compared to $\delta M_2=0.01$, which suggests a correlation between non-Kerr deviation and vertical shadow size.
However, the mass and spin of the central object are not known in priori, so the absolute size of the shadow cannot be used as a convincing observable for the non-Kerrness of the object.
 Because of this reason, instead of focusing on the comparison with a particular Kerr BH shadow, we need to consider the difference in ``shape" between the shadow of the central object and a Kerr shadow. We will define a  measure for the area of the mismatched region between this compact-object shadow and Kerr shadows with a range of mass and spin parameters. For this purpose, we apply a semi-analytical approach for computing Kerr shadows which is computationally less expensive than evolving the null geodesics in the ray-tracing method.
\subsection{Semi-Analytical Kerr Shadow}\label{sec:4B}
The Kerr spacetime is described by the metric below in the \textit{Boyer-Lindquist} coordinates:
\begin{eqnarray}    
    ds^2 &=&- \bigg( 1 - \frac{2 M r}{\Sigma} \bigg) \  \mathrm{d}t^2 - \frac{4 a M r \sin^2 \theta}{\Sigma} \mathrm{d}t \ \! \mathrm{d}\phi  + \frac{\Sigma}{\Delta} \mathrm{d}r^2 \nonumber \\
    & & \hspace{-0.25cm} + \Sigma \mathrm{d}\theta^2 + \bigg( r^2 + a^2 + \frac{2 a^2 M r \sin^2 \theta}{\Sigma} \bigg) \sin^2\theta \ \!  \mathrm{d}\phi^2,
    \label{BL Kerr}
\end{eqnarray}
with $\Sigma$ and $\Delta$ being:
\begin{eqnarray}
    \Sigma &\equiv& r^2 + a^2 \ \cos^2\theta , \\
    \Delta &\equiv& r^2 \ -2Mr + a^2.
\end{eqnarray}
where $M$ and $a$ are the black hole's mass and the spin parameter respectively.
The Kerr shadow can be computed semi-analytically following the approach discussed in~\cite{Chandrasekhar:1985kt, Gralla_2018}. In general,  spherical null orbits in Kerr spacetime can be characterized by solving the equations
\begin{equation}
    V_r(r) = 0, \quad V_r'(r) = 0
    \label{eqn:potentials}
\end{equation}
simultaneously. Here prime ($'$) denotes the derivative with respect to $r$.  $V(r)$   is the radial potential of the Kerr metric, in a way that we can write the radial geodesics equation as follows \cite{Bardeen}:
\begin{equation}
\begin{split}
    \Sigma \dot{r} &= \pm (V_r)^{1/2} \,, \\
    &=  [ E \ (r^2 + a^2) - L \ a]^2 - \Delta \ [ (\ L - a \ E)^2 + Q^2] 
\end{split}
\label{eqn: radial potential}
\end{equation}
and the dot is the derivative with respect to the affine parameter. $E$, $L$,
and $Q$ are energy, angular momentum, and Carter constant of the photon, respectively. Solving the equation \ref{eqn:potentials}, we find the radius of unstable spherical null orbits ($r_{ph} (E, L, Q)$). The aim here is to find sets of these constants of motions that could distinguish the photons passing the $r_{ph}$ and falling into the horizon from those that run to infinity. Additionally, the null rays in the Kerr spacetime can be described by the two following independent ratios: 
\be
\lambda = \frac{L}{E}, \quad q = \frac{Q}{E^2},
\label{impact params}
\ee
$\lambda$ and $q$ are characteristics of a null ray direction as seen by a distant observer, or so-called impact parameters \cite{Chandrasekhar:1985kt}. Hence, instead of using the three constants of motion, we can deal with these two impact parameters to classify the null geodesics. The first step to calculate the shadow border is to use the translation of the impact parameters into coordinates in the camera frame, using the notations in~\cite{Bardeen:1973tla,Gralla_2018}:
\begin{equation}
x_c  =  - \frac{\lambda}{sin \theta_O}, \quad  y_c = \pm \sqrt{q + a^2 cos^2 \theta_O - \lambda^2 cot^2 \theta_O} 
\label{eqn:coordinates}
\end{equation}
The $\pm$ determines the region above and below the equatorial plane respectively. 
Now, using the solution of \ref{eqn:potentials}, we find the radius of unstable spherical null orbits ($r_{ph} (E, L, Q)$), accordingly we can inverse that to write both impact parameters in terms of $r_{ph}$:
\begin{equation}
\begin{split}
    \lambda &= -\frac{r_{ph}^2 (r_{ph} - 3 M) + a^2 (r_{ph} + M)}{a (r_{ph} - M)}\,, \\
    q &= \frac{r_{ph}^3 (4 a^2 M - r_{ph} (r_{ph} - 3M)^2)}{a^2 (r_{ph} - M)^2}  \,,
\end{split}
\label{impact params2}
\end{equation}
This means that for any set of impact parameters $[\lambda, q]$, we would have two corresponding coordinates $[x_c, y_{c \ \pm}]$ in the camera frame. 
Photons on an unstable spherical orbit would inevitably pass the equatorial plane, and Carter constant by definition $ -Q = p^2_\theta + [ L^2 \csc^2 \theta - a^2 E^2 ]\cos^2 \theta- $ would be positive for $\theta = \pi/2$. As a result, a photon passing the equatorial plane has a positive Carter constant, $Q>0$. A zero Carter constant corresponds to orbits in the equatorial plane with $p_{\theta} = 0$. Consequently, for spherical orbits, considering the Eq. \ref{impact params}, $q$ should be $q \geq 0 $. The solution to $q=0$ gives us the radii of equatorial unstable circular orbits for photons co-rotating, and counter-rotating the BH's spin, $ r_{ph}^{\pm} = 2 M \big[ 1 + \cos \big(2/3 \ \cos^{-1}(\mp a/M)\big) \big] $. To extract the allowed values of $[x_c, y_c]$ representing the shadow border, we inserted different values for $r_{ph}$ from the interval of $[r_{ph}^-,r_{ph}^+]$ into Eq.~\ref{impact params2}, which results only in non-negative values of $q$, and an interval for $\lambda$. Additionally, for a non-equatorial observer $\theta_O \neq \pi/2$, we applied the condition of $y_c$ being real as well. 
Comparison between a Kerr shadow computed semi-analytically and using the numerical method in Sec.~\ref{sec:4A} is presented Fig.~\ref{fig:shadows} and Fig.~\ref{fig:shadows_rth} which show that they are fairly consistent.

\subsection{Shadow degeneracy}
The shape of the shadow of a Kerr black hole depends on its spin, mass, and the inclination angle $\theta_O$ of the observer's location with respect to the black hole's coordinate system. Since these parameters are not known in priori, for a given measure of mismatch, we should sample all the Kerr shadows in the  $(a\,, M\,, \theta_O)$ parameter space and determine the minimal mismatch with the measured data. If this mismatch is larger than the sensitivity limit of the detector, then one can claim a positive evidence of Kerr deviation.
For this purpose, we define the ``mismatch" between two shadows as   the area of non-overlapping parts of these two shadows, after co-centering them at the origin of the camera's reference frame. This allows us to measure the accumulated deviations along the edge of the shadows, summarizing them into a single variable. Mathematically, this mismatch quantity can be defined as $\text{Mismatch} \equiv \int_0^{2\pi} |\delta r_c| d \phi_c$, which is essentially the magnitude of the area in the $\delta r_c$ vs $\phi_c$ plot, where $r_c=\lb x_c^2+y_c^2\rb^{1/2}$.

\begin{figure}[t]
    \centering
\subfigure{
    \includegraphics[width = 8cm]{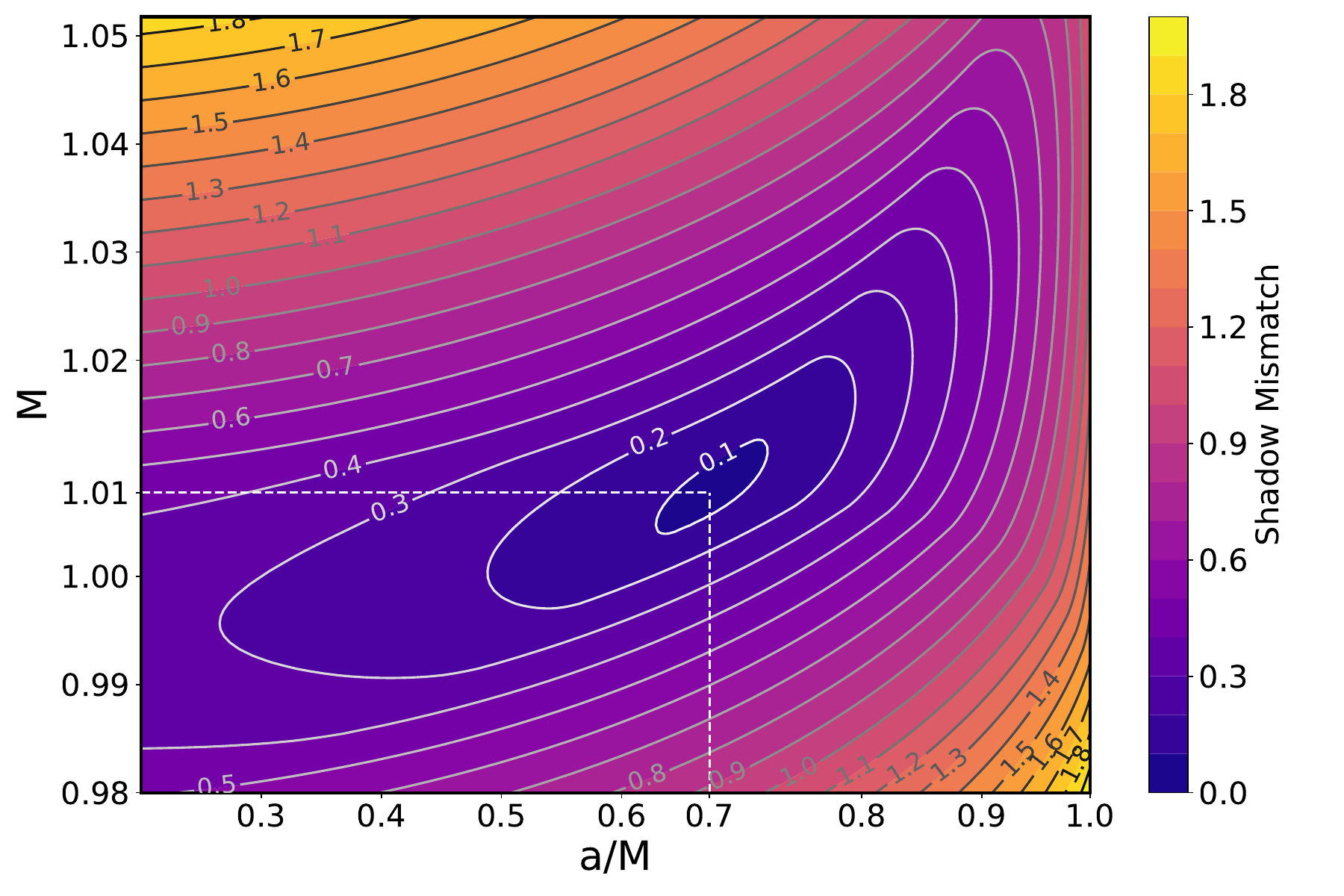}
    \label{fig:mis-th90}
}
\subfigure{
    \includegraphics[width = 8cm]{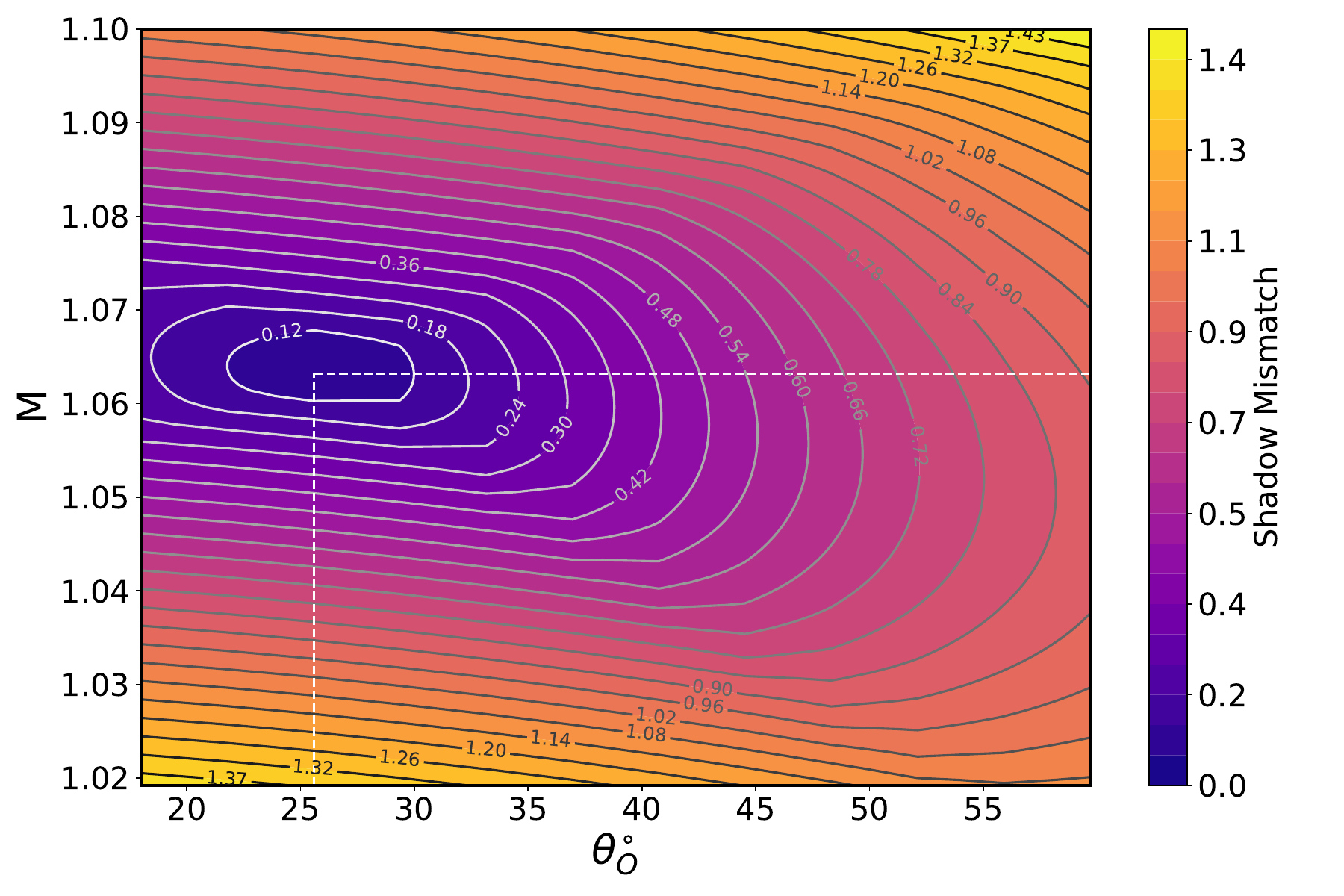}
    \label{fig:mis-th37.8}
}
\caption{The mismatch contour plots in spin vs. mass space. Injected shadow is the spacetime with $ \delta M_2 = 0.08,\, a=0.3,\, M = 1$. \textcolor{black}{The plot at the top shows that the minimum mismatch of a Kerr shadow in the equatorial point of view with the injected shadow happens for $a = 0.67$ and $M = 1.0077$ with a relative deviation of 0.35\%. The plot at the bottom shows the mismatch in the $M-\theta_{O}$ plane for fixed spin of $a=0.99$. In this case, the minimum mismatch happens for $\theta_O = 25.58^{\circ}$, $M = 1.0632$, and the relative deviation is 0.17\%.}}
    \label{fig:mis2d}
\end{figure}

We calculate the mismatch parameter between the shadow of Kerr spacetime and the spacetime with $\delta M_2 = 0.08$ in two different scenarios. Firstly, we fix the observer's inclination angle $\theta_O = 90^{\circ}$ (Fig.~\ref{fig:mis2d} top panel) for various mass and spin values within the range $M \in [0.98, 1.1]$, and $a \in [0.2, 0.99]$, with the $M=1$ in our simulation results. \textcolor{black}{Secondly, we allow $\theta_O$ to vary (Fig.~\ref{fig:mis2d} lower panel). The minimum mismatch parameter with a Kerr shadow for the $\theta_O = 90^{\circ}$ case in such a parameter regime has been found to have a relative deviation \footnote{Relative deviation is defined as the mismatch parameter over the total area of the simulated shadow} of 0.35\% which is realized at $ a = 0.67$ and $M = 1.0077 $. On the other hand, by allowing the variation of the angle $\theta_O \in [\pi/10, \pi/2]$, the minimum mismatch of 0.17\% has been found for $\theta_O = 25.58^{\circ}$, $ a = 0.99$ and $M = 1.0632 $. 
It is worth mentioning that the relative deviation of the shadows from the Kerr of the same spin and mass values is $0.92\%$ and $0.11\%$ for $\delta M_2 = 0.08$, and $\delta M_2 = 0.01$ respectively. This means that the mismatch analysis for the case of $\delta M_2 = 0.01$ can be approximated to be linearly scaled.}

\begin{figure}[t]
    \centering
    \includegraphics[width = 8.5cm]{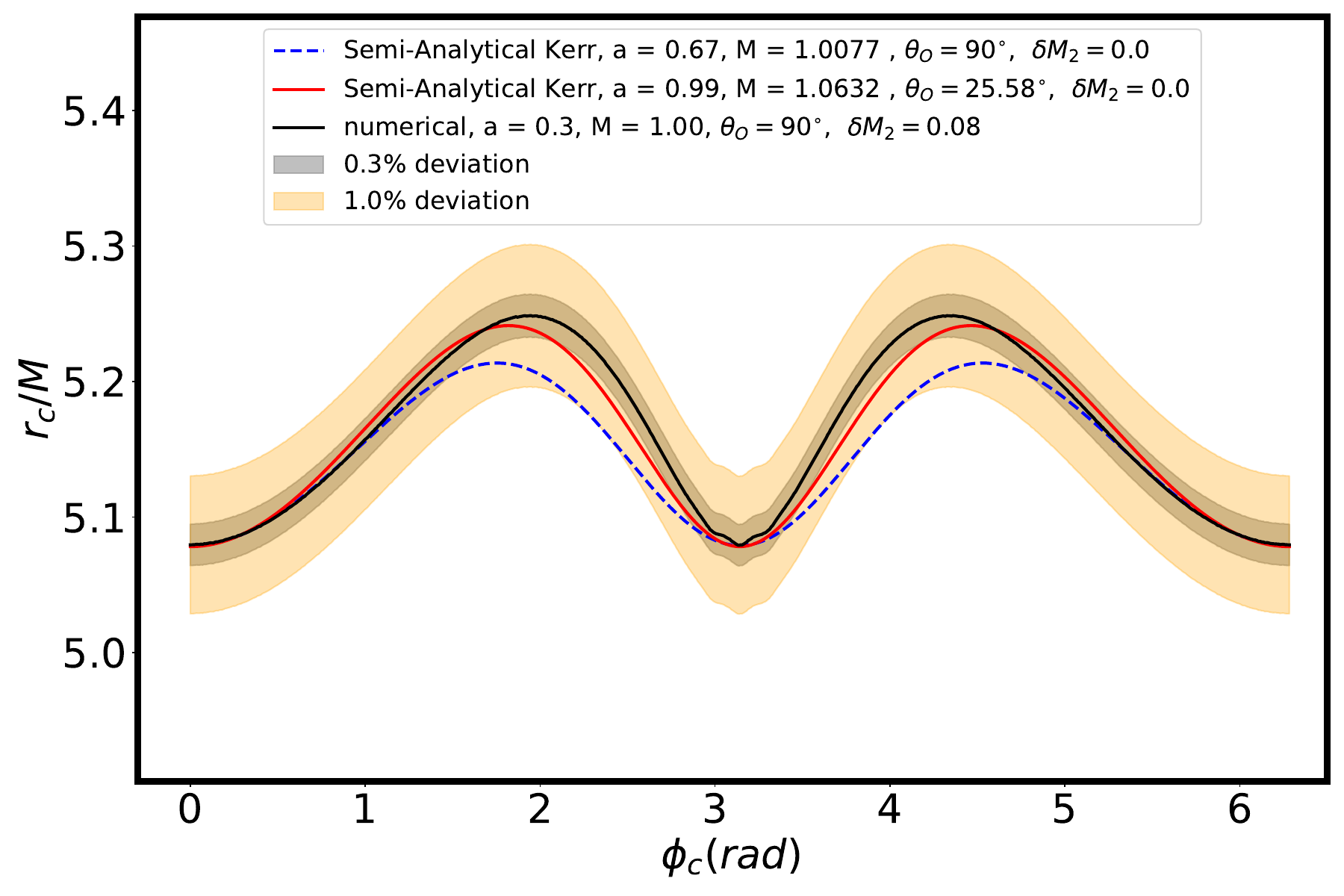}
    \caption{\textcolor{black}{In this figure we show the shadow border ($r_C$) corresponding to two minimum mismatch cases discussed in Fig.~\ref{fig:mis2d}. Two highlighted regions reflect $ 0.3\% $ (gray), and $1.0\%$ (yellow) uniform deviation from the central object of $\delta M_2=0.08$ case. This shows the detection resolution needed for distinguishing the Kerr with different mass and spin values from our multipole spacetime.}}
    \label{fig:deviation}
\end{figure}

Fundamental physics tests with black hole imaging may be limited by astrophysical uncertainties in modeling the accretion flows and their emissions, so it is important to identify observables that are less susceptible to the influence of astrophysical uncertainties \cite{Gralla_2021}. The light-ring/critical-curve signatures of black holes may serve as a viable option \cite{Gralla_2020}.  
For the sample system considered, in order to detect $\delta M_2$ to the level of $0.08 M^3$, the detector has to be able to resolve shadow mismatch at the level of \textcolor{black}{$\approx 0.3\%$ (See Fig.~\ref{fig:deviation})}. This sensitivity requirement may only be realized with Earth-space or space-based VLBI \cite{Gralla_2020, lazio2020space}. On the other hand, because of the degeneracy between parameters, if a central object indeed has nonzero $\delta M_2$ but its shadow is fitted with a Kerr black hole template, the inferred values of the spin may be severely biased. 

\section{EMRI Waveform}\label{sec:5}
Extreme mass-ratio inspirals (EMRIs) generally comprise a massive black hole and a stellar-mass compact object. The mass ratio,
denoted as $q = \frac{\mu}{M}$ (with $\mu$ being the smaller mass), typically falls within the range of $q = 10^{-6} - 10^{-4}$. The evolution timescale of an EMRI approximately scales inversely with the mass ratio  ($q^{-1}$) as it is driven by the gravitational radiation reaction. Consequently, there is a clear separation of timescales between the orbital timescale and the evolution timescale of orbital energies. The smaller compact objects typically spend a significant number of orbits ($10^4 -10^5$) in the detector's band before finally plunging into the more massive object. This extended period of observation allows the accumulation of small changes in the spacetime's structure or environmental effects over many cycles to be amplified.

Therefore EMRIs are ideal probes for the spacetime of the central object. A small variation of the multipole moments generally leads to a metric different from Kerr, so that the EMRI evolution within the underlying spacetime and the corresponding waveform are also modified. Similar studies of the effect of deviation in quadrupole moment on the orbits around the massive compact objects have been done in \cite{Collins_2004} using their \textit{``Bumpy Black Hole"} models. However, we are using the metric computed in the first part of the paper which fully describes the entire spacetime even in the strong regime. In this work, we consider an EMRI system comprising a massive central object associated with the beyond-Kerr spacetime and a compact object in the form of a stellar mass point particle that inspirals towards it. By comparing the accumulated phase of the gravitational wave emitted by this system to that of the Kerr spacetime, we can quantify the dephasing between these two spacetimes. This dephasing serves as an observable that can be used to probe the spacetime multiples.

\begin{figure}[t]
\centering
\subfigure{
\includegraphics[width = 8cm]{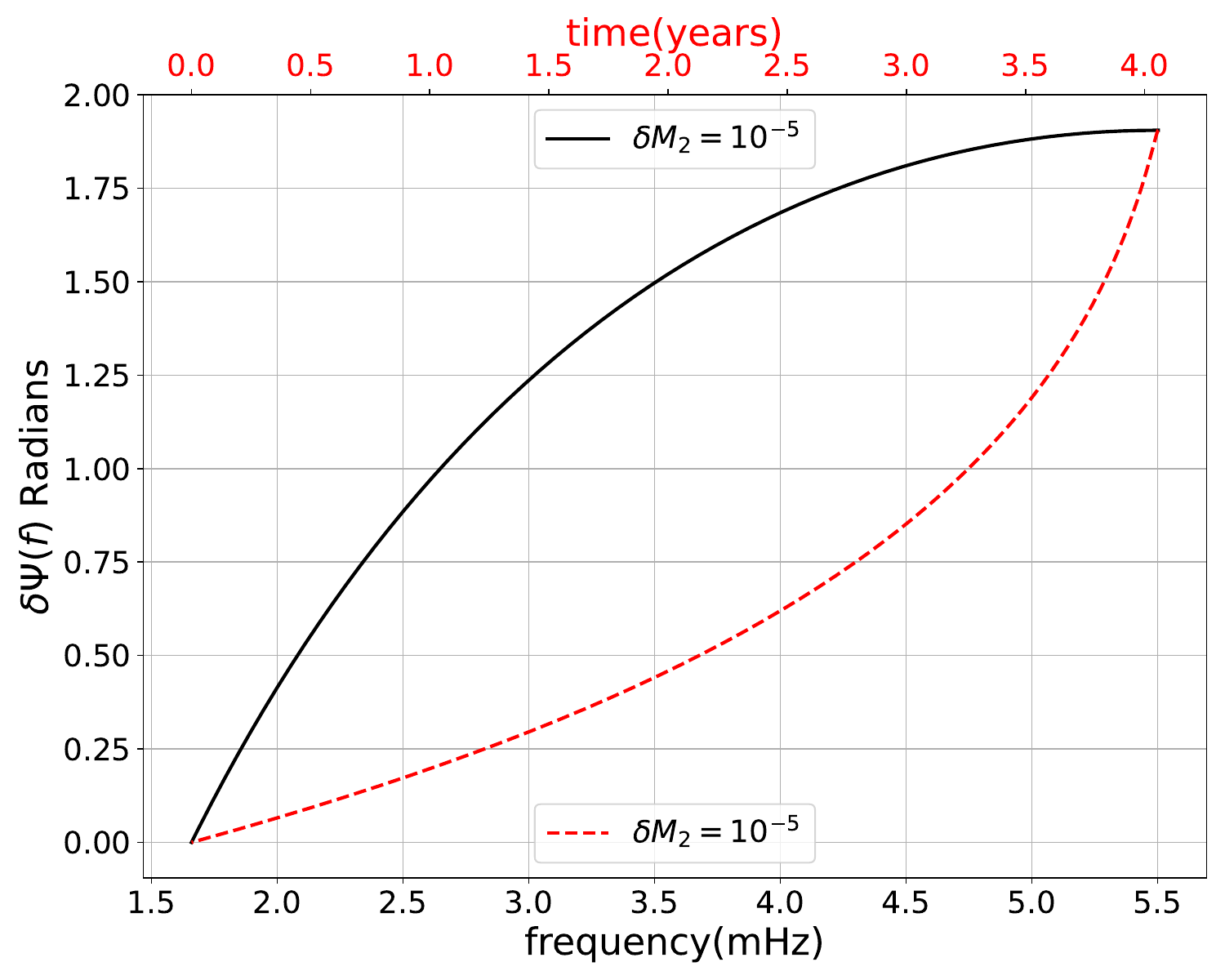}
\label{fig:mis-th90}
}
\caption{\textcolor{black}{Accumulated phase difference (dephasing) of the gravitational wave as a function of both time and frequency, during 4 years of observation time. We have considered a point source of mass $\mu = 10 M_{\odot}$ in an equatorial quasi-circular orbit around the central object of mass $M = 10^6 M_{\odot}$, spin parameter of $a=0.3$ with the deviation in quadrupole moment $\delta M_2 = 10^{-5}$. The starting, and ending orbital radii are chosen to be $11.43 M$, and $5.0 M$ respectively, such that the secondary object plunges within the 4 years of observation. }  }
    \label{fig:phase_shift}
\end{figure}

We have chosen a sample  EMRI system with a secondary compact object of mass $\mu = 10 M_{\odot}$ moving in an equatorial quasi-circular orbit, adiabatically inspiralling towards the central body of mass $M = 10^6 M_{\odot}$ and spin $a = 0.3$. The initial radial distance is set to be $\blue{r_i = 11.43 M}$ and the final radius (of consideration) $\blue{r_f = 5.0 M}$ and consequently 
the orbit starts from a gravitational wave frequency of \blue{$f_{GW} \approx 1.6\,mHz$}, and ends before it reaches the Kerr ISCO ($r_{ISCO} = 4.978 M$ for Kerr black hole of spin a=0.3). Therefore we believe the $\blue{r_f = 5 M}$ would be a safe choice for our analysis making sure that the adiabatic approximation would still hold, and it is a typical choice to be compared with an EMRI with a non-rotating central black hole. The entire orbit would then correspond to the GW frequency interval of \blue{$f_{GW} \in [1.5,5.5]\,mHz$}, which lies within the LISA frequency sensitivity band \cite{Robson_2019}.

In the adiabatic approximation, the secondary object moves along the instantaneous geodesics of the background spacetime on a time scale much shorter compared to the radiation reaction timescale \cite{Hinderer_2008}. Following this argument and considering the Fourier transform of the waveform under the stationary phase approximation \cite{Droz_1999}, the gravitational waveform may be written as
\begin{align}
 h(f) \sim A(f) e^{i \psi(f)}\,, 
\end{align}
 where the total phase may be computed by using the energy of the secondary body and the energy loss rate  \cite{Vines_2011} :
\begin{equation}
    \frac{d^2 \psi}{d \Omega ^2} = \frac{2 E'(\Omega)}{\Dot{E}}\,.
    \label{eq:d2psi}
\end{equation}
Here the prime denotes the derivative with respect to $\Omega$, and  $\Dot{E}$ is given by  $\frac{d E}{dt} = -\mathcal{F}$, with $\mathcal{F}$ being the total energy flux radiated to infinity and down toward the central object (to the Horizon in case of Kerr). Equation Eq.~\eqref{eq:d2psi} can be easily rearranged in terms of the gravitational wave frequency $f$ (of the $22$ mode) rather than the orbital angular frequency $\Omega$ by using $\Omega = \pi f$. Throughout the computation process, we are using the geometrized units $G = c = 1$, and we set $M = 1$ which is the total mass of the spacetime. At the end to make the plots, we recover the units to report the results in SI units. 

Because the spacetime of the central object is only weakly perturbed from Kerr, we write down the bakcground metric as
\begin{equation}
    g_0 = g_{K} + \epsilon h\,.
    \label{metric_perturb2}
\end{equation}
where we split the background metric $g_0$ into two parts, the Kerr metric $g_{K}$ plus a modification $\epsilon h$ due to the deviation in the quadrupole moment ($\delta M_2 \neq 0$). Here $\epsilon$ is a book-keeping index. Our main goal is to compute the accumulated phase difference $\delta \psi = \psi_{\delta M_2} - \psi_K$ as a function of $\Omega$ for a fixed value of $\delta M_2$. To obtain $\delta \psi$, we could use the approximation below for the \textit{right-hand side} of Eq.~\eqref{eq:d2psi}, up to linear order in $\epsilon$:
\begin{equation}
     \frac{d^2 \delta\psi}{d \Omega ^2} = \delta \left (\frac{2 E'(\Omega)}{\Dot{E}} \right ) \approx  \left (\frac{2 \delta E'(\Omega)}{\Dot{E}} \right )-\frac{2 E'(\Omega) \delta \Dot{E}}{\Dot{E}^2}\,,
    \label{eq:deltad2psi} 
\end{equation}
where
\begin{eqnarray}
     \delta E &=& E - E_K \,, \\
     \delta E'(\Omega) &\equiv& \frac{ \delta dE}{d \Omega}  =  \frac{ dE}{d \Omega} - \frac{ dE_K}{d \Omega} \,.
     \label{eq:deltaEprime linear}
\end{eqnarray}

Before proceeding further, it is important to note that in this work we are only considering the waveform modulation due to $\delta E$ in equation Eq.~\eqref{eq:deltad2psi}. In general $\delta \dot E$ is not zero and should have modifications of the same order in both the mass-ratio and $\epsilon$, although the frequency dependence may be different. However, it requires developing a modified Teukolsky equation in order to obtain the modified energy flux \cite{Pan:2023wau}, which is beyond the scope of this work. By disregarding the terms associated with $\delta \dot E$, we can still obtain a result that provides a reasonably accurate estimate of the order of magnitude for $\delta \psi$, albeit without incorporating those modifications.

On the other hand, let us consider a stationary axisymmetric spacetime in \textit{Boyer-Lindquist} coordinates with line elements of:
\begin{equation}
    ds^2 = g_{tt} dt^2 + 2 g_{t\phi} dt d\phi + g_{rr}dr^2 + g_{\theta\theta} d \theta^2 + g_{\phi\phi}d\phi^2\,.
\end{equation}
The conserved energy of a test particle orbiting around the central body is: 
\begin{equation}
    \frac{E(r, \delta M_2)}{\mu} = - g_{tt}\frac{dt}{d\tau}  - g_{t\phi}\frac{d\phi}{d\tau} \,,
    \label{eq:energy}
\end{equation}
assuming that the test particle moves along an equatorial, circular orbit ($\theta = \frac{\pi}{2}$, $\frac{dr}{d\tau} = 0$, and $\frac{d \theta}{d \tau} = 0$). The angular velocity can be calculated as follows:
\begin{equation}
    \Omega(r, \delta M_2) = \frac{d \phi}{dt} = \frac{ -g_{t\phi,r} + \sqrt{ g_{t\phi,r}^2 - g_{tt,r} g_{\phi\phi,r} }  }{g_{\phi\phi,r}}
    \label{eq:omega}
\end{equation}
for which all the relevant metric components are functions of the spin parameter, $a$, the radial distance from the center, $r$, the polar angle $\theta$ and $\delta M_2$. The above formula is useful as we can compute  $E'(\Omega) = \frac{dE}{d\Omega}$ using the metric data in Eq.~\eqref{eq:energy}, Eq.~\eqref{eq:omega} and the chain rule:
\begin{equation}
    \frac{dE}{d\Omega} = \frac{dE}{dr} \left (\frac{d\Omega}{dr} \right )^{-1}\,.
\end{equation}

Since the metric for this spacetime is expressed as power-law expansions with a large number of terms for each component, it is not straightforward to write down a compact analytical expression for $dE/d\Omega$, which is computed numerically. Operationally we compute  $E'(\Omega(r))$ and  $\Omega(r)$ on the same grids of $r$ that cover the relevant parameter range, and numerically interpolate $E'$ over $\Omega$, such that the right-hand side of Eq.~\eqref{eq:deltaEprime linear} can be evaluated.  
In addition, to compute the accumulated phase using Eqs.~\eqref{eq:deltad2psi}, we have numerically computed the Teukolsky flux including $(l,m)$ modes up to $l_{max} = 10$, on the same grid points of $r$ (and therefore the same grid of $\Omega$) for a point particle orbiting a Kerr black hole of spin $a=0.3$. Having these interpolated numerical functions, we perform the numerical integration twice on the right-hand side of Eq.~\eqref{eq:deltad2psi} to obtain $\delta \psi(\Omega)$.

Although for the sample problem, we are assuming a spacetime with $\delta M_2 = 0.08$, the linear approximation in $\epsilon$ for the $\delta \psi$ enables us to compute the waveform modulation corresponding to other small $\delta M_2$, as  $\delta \psi \propto \delta M_2$. To validate the accuracy of our linear approximation, we have used the exact metric data for two other different values of $\delta M_2$, and compared the resulting values of the final $\delta \psi$ with those obtained from the linear approximation in the table \ref{tab:comparison}, while assuming a fixed value of mass-ratio $q = 10^{-5}$. According to the table \ref{tab:comparison}, the linear approximation is valid to a good accuracy. 

\textcolor{black}{Fig.~\ref{fig:phase_shift}, demonstrates the accumulated dephasing $\delta \psi(f)$ as a function of actual gravitational wave frequency for a sample case of $\delta M_2 = 10^{-5}$. This value of $\delta M_2$ has been chosen only to require that the minimum phase shift is still above a conservative detection threshold of $1 {\rm\,rad}$ \cite{Bonga_2019}. Since LISA has at least a four-year observation window, we consider the last 4 years of the EMRI evolution before the plunge in the Fig.~\ref{fig:phase_shift}.
}
\begin{table}[h]
    \centering
    \begin{tabular}{|c|c|c|c|}
    \hline
        $\delta M_2$ & 0.08 & 0.16 & 0.0008\\ \hline
        Relative Error &  \textcolor{black}{0.036\%} & \textcolor{black}{0.59\%}& \textcolor{black}{0.5\%} \\ \hline
         $\delta \psi / \delta \psi_{0.08}$ & \textcolor{black}{$1.0 \  (\pm 10^{-3})$} & \textcolor{black}{$2.0 \ (\pm 10^{-3})$}  & \textcolor{black}{$10^{-2} \  (\pm 10^{-5})$} \\ \hline  
    \end{tabular}
    \caption{In this table, we present the accuracy of the linear approximation. In the second row, the relative error is the relative difference between the actual phase difference for the corresponding value of $\delta M_2$ and the scaled version of the linear phase difference. The third row also shows the ratio between phase shift corresponding to each $\delta M_2$ and the main case of $\delta M_2 = 0.08$, which actually shows how linear they are ( up to $\delta M_2=0.16$). Evidently, the linear approximation is applicable to our choice of parameters where $\delta M_2<0.09$.}
    \label{tab:comparison}
\end{table}

Since EMRI evolution generally follows  the adiabatic approximation  except at the plunge phase, the gravitional wave phase can be expressed as expansions of $1/q$, with the leading order term being $\delta \psi \propto q^{-1}$  \cite{Hinderer_2008}. However, if the period of observation is smaller than the period of the EMRI staying in band, the accumulated dephasing is mainly limited by the observational period instead of the the radiation reaction timescale.
In this case, the dephasing is approximately
\begin{align}
\delta \psi \sim  \mathcal{O}(1)\left ( \frac{\delta M_2}{10^{-5}M^3}\right )  \,.
\end{align}
It is evident that EMRI systems are superior probes of the spacetime multipole moments compared to direct observation of the critical curve of the central object by VLBIs.

\section{Conclusion}\label{sec:6}
BH mimickers may support  spacetimes with arbitrary multipolar structures. The field multipole moments at large distances are limited by the source's properties, such as its size and motion, which contribute to the source multipole moments. Our study explores how field multipole moments should scale with the size of a source in the strong-field relativistic limit, particularly for the cases with large  moments. We find that the source's size should be smaller than the radius of convergence of metric components expressed as a Taylor series expansion in the inverse radial distance. Determining the radius of convergence requires a metric accurate up to a sufficiently high order in the inverse radial distance. We have implemented the Ernst formalism in an axistationary spacetime for that purpose. Our findings indicate that for sufficiently large Geroch-Hansen multipole moments in a static axisymmetric spacetime, the dependence of such moments on source length scale is remarkably different from Newtonian expectations. The characteristic size of the source $L$ scales with the scalar multipole moment $M_n$ as $M^{1/{(n+1)}}$ instead of the traditional Newtonian scaling of $M_n^{1/n}$. This implies that a source of a smaller size than the Newtonian estimation can create an equally large multipole moment.

In order to test the ``Kerrness" of a spacetime, it is interesting to study those with small deviations from the Kerr moments.
In particular, we have considered the case with a small deviation from a Kerr quadrupole moment and have semi-quantitatively estimated the relation between minimal size and the non-Kerr quadrupole moment using an upper bound test on the curvature. 
There is a correlation between the non-Kerr quadrupole moment and the size of the source when it comes to the pattern exhibited by the radius of  curvature threshold, assuming the location of the curvature threshold is directly related to the minimal  size of the source.
For example, for a positive non-Kerr quadrupole deviation, we find that a larger size for the object (in terms of coordinate values) corresponds to a bigger non-Kerr quadrupole modification when the spin is kept fixed.

In order to measure non-Kerr deviations in the quadrupole moment to observations, we have looked into shadows of compact objects observable by EHT. We have implemented the backward ray-tracing method to compute the shadows of these compact objects. In order to probe the difference between Kerr shadows and a shadow created by a compact object with quadrupole deviation, we have computed the mismatch in the area of shadows. As a sample spacetime with mass set to be unity, spin parameter set to be $a =0.3$, and a quadrupole moment deviating from the Kerr value by approximately 88\%  in magnitude, our analyses show a minimum mismatch of \textcolor{black}{0.17\% with a Kerr shadow of a BH with mass $M=1.0632$ and a spin of $a=0.99$}. This means that the parameter degeneracy seriously limits our ability to measure the spacetime multipole moments with EHT observations.

Compared to the challenges with  EHT observations, we have also examined the possibility of measuring GWs from EMRIs to probe the non-Kerr deviations. EMRIs are unique in the sense that the time spent in the inspiral phase is much longer than the orbital timescales, and  future space-based telescopes such as  LISA will be able to harness such opportunity to measure the black hole spacetime. For example, considering a central object of mass of $10^6$ solar mass with spin being $a=0.3$, our analyses show that such observations will be sensitive to the deviation from Kerr quadrupole moments  that is at least $0.01\%$ in the relative magnitude compared to Kerr.

We conclude by pointing to several possible future extensions of our work. Firstly, one can investigate the minimal size conjecture in General Relativity  without assuming the axisymmetry. A possible approach is to compute the conformal metric, keeping desirable multipoles in a set of normal coordinates around spatial infinity as outlined in Ref.~\cite{Herberthson:2009ze}, and then transforming back to physical spacetime. However, computing metrics to a sufficiently high  order to determine the convergence radius may be time-consuming without assuming certain symmetries. Secondly, the analysis using  EMRI systems to measure the spacetime of a compact object requires the calculation  of modified GW energy flux in such perturbed spacetime. Such a task is nontrivial as it calls for a proper calculation with modified Teukolsky equations. The EMRI motion near the resonant regime also has to be accounted for \cite{Pan:2023wau}. Finally, we can include non-Kerr deviations to other multipole moments in addition to the quadrupole to investigate  how they affect the conclusions of our analyses.

\acknowledgements

We would like to thank Hung-Yi Pu for providing valuable guidance on their publicly available ray-tracing code, ``Odyssey", and for taking the time to address our related questions. We thank Beatrice Bonga and Eric Poisson for insightful discussions on the Weyl-Papapetrou metric and multipole moments at the early stage of this work. We Also thank Zhen Pan for his initial input on computing EMRI frequency shift for multipole spacetime metric. We are grateful to Luciano Combi for his expert advice on addressing the numerical challenges encountered during this project. Additionally, we would like to acknowledge Samuel Gralla for his insightful advice and discussions on the critical curve of black holes, conducted during his visit to the Perimeter Institute. We are supported by the Natural Sciences and
Engineering Research Council of Canada and in part by
Perimeter Institute for Theoretical Physics. Research at
Perimeter Institute is supported in part by the Government
of Canada through the Department of Innovation, Science
and Economic Development Canada and by the Province of
Ontario through the Ministry of Colleges and Universities. We have performed our numerical simulations on the ``Symmetry" HPC at the Perimeter Institute.
\bibliography{ms}

\end{document}